\documentclass[english]{revtex4}
\usepackage[T1]{fontenc}
\usepackage[latin9]{inputenc}
\usepackage{amsmath}
\usepackage{amssymb}
\usepackage{wasysym}

\makeatletter
\@ifundefined{textcolor}{}
{%
 \definecolor{BLACK}{gray}{0}
 \definecolor{WHITE}{gray}{1}
 \definecolor{RED}{rgb}{1,0,0}
 \definecolor{GREEN}{rgb}{0,1,0}
 \definecolor{BLUE}{rgb}{0,0,1}
 \definecolor{CYAN}{cmyk}{1,0,0,0}
 \definecolor{MAGENTA}{cmyk}{0,1,0,0}
 \definecolor{YELLOW}{cmyk}{0,0,1,0}
}

\makeatother

\usepackage{babel}
\begin{document}
\title{The Ricci Flow and the Early Universe}
\author{M.J.Luo}
\address{Department of Physics, Jiangsu University, Zhenjiang 212013, People's
Republic of China}
\email{mjluo@ujs.edu.cn}

\begin{abstract}
A framework of quantum spacetime reference frame is proposed and reviewed,
in which the quantum spacetime at the Gaussian approximation is deformed
by the Ricci flow. At sufficient large scale, the Ricci flow not only
smooths out local small irregularities making the universe a homogeneous
and isotropic Friedmann-Robertson-Walker metric, but also develops
a local singularity at the physical-time origin. Due to the phenomenological
suppression of the non-Gaussian primordial perturbations, we assume
the validity of the Ricci flow applying to the high curvature region
near the local singularity of the early universe. The no-local-collapsing
theorem of Perelman ensures the existence of a canonical neighborhood
around the large curvature pinching point, which resembles a gradient
shrinking Ricci soliton (GSRS) solution of the Ricci flow. Without
any inflaton field, the GSRS naturally reproduces an exact inflationary
deSitter universe near the singularity at the leading order. Without
any rolling-down behavior of inflaton, the deviation from exact deSitter
described by the ``slow roll parameters'' can be calculated by a
small deviation from the singular flow-time via the Ricci flow, and
the primordial perturbations can also be studied on the GSRS background,
the power spectrum of the scalar perturbation agrees with present
observations, and the one of the tensor perturbation is predicted
too small to be detectable than the standard inflation. The previous
treatment of the cosmological constant and the effective gravity are
also briefly reviewed in the framework. So we argue that the Ricci
flow provides us a possible unified view and treatment of the late
epoch accelerating expansion and early epoch inflation of the universe
without introducing dark energy or inflaton (dark energy of the second
kind).
\end{abstract}
\maketitle

\section{Introduction}

Issues concerning the origin of the universe and its primordial structure
are among the deepest and most fundamental in physics. The textbook
standard inflation theory, and a wide class of inflation paradigm
or alternative theories are motivated by solving the fundamental problems
of the big-bang theory of the universe at the early epoch, for instance,
the homogeneous problem, the horizon problem and the flatness problem
etc. On the one hand, the standard prediction from the simplest inflationary
picture is extremely consistent with recent observation from the Cosmic
Microwave Background (CMB) radiation \citep{Planck:2018vyg}, on the
other hand, the inflationary paradigm is considered having its own
conceptual problems (e.g. see \citep{Ijjas:2014nta,BRANDENBERGER2014109}
and references therein), certainly the statement itself is also controversial
\citep{Chowdhury:2019otk}. At the fundamental level, in the absence
of quantum description of gravity and spacetime, it is difficult to
know which of the problems of inflation are more serious and correct
to ask in order to get a better description of physics near the origin
of the universe. However, it is no controversial that the basis of
the inflation cosmology does not yet have the status of an established
theory, the mechanism of inflation is based on the speculative inflaton
in the early epoch (another unknown form of dark energy in the late
epoch), the inflation theory (and other alternative theories) of the
early universe are not yet fully quantum, it is based on a semi-classical
combination of the classical general relativity and the quantum theory.

In the previous literature \citep{Luo2014The,Luo2015Dark,Luo:2015pca,Luo:2019iby,Luo:2021zpi,Luo:2022goc,Luo:2022statistics,Luo:2022ywl},
a framework applying the quantum principle to a spacetime reference
frame and gravity is proposed, in which the equivalence principle
and the general covariance are generalized to the quantum level. It
shows that the quantum behavior of the spacetime to a large extent
is embodied in its Ricci flow. And the renormalizability of the theory
is mathematically related to the uniformization of the spacetime to
be proved by the Ricci flow approach. The Ricci flow was firstly introduced
in 1980s by Friedan in $d=2+\epsilon$ non-linear sigma model \citep{friedan1980nonlinear,Friedan1980}
and also independently invented by Hamilton in mathematics \citep{Hamilton1982Three,hamilton1986four}.
From the mathematical side, the main motivation of the Ricci flow
is to classify manifolds, especially, to prove the Poincare conjecture.
Hamilton's program is to use the Ricci flow as a useful tool to gradually
deform a manifold into a ``simpler and nicer'' manifold whose topology
can be readily recognized. But unfortunately, the program met some
difficulties in treating generic initial conditions, because in general
the flow may develop local singularities. A general realization of
the program is achieved by Perelman at around 2003 \citep{perelman2002entropy,perelman2003ricci,perelman307245finite},
who introduced several monotonic functionals to successfully deal
with the local singularities. The monotonicity of the functionals
are applied to prove the ``no-local-collapsing theorem'', and to
prove the existence of a regular subset of the manifolds around the
local singularity called ``canonical neighborhood structure''. It
is shown that the local singularity or the high curvature region around
it belongs to finite number of ``singularity models'' resembling
the ``gradient shrinking Ricci soliton'' (GSRS) configurations.
A surgery can be performed in the canonical neighborhood around the
singularity, and then the Ricci flow is able to continue, the breakthrough
of Perelman finally removed the stumbling block in Hamilton's program.
The existence of the canonical neighborhood structure around a local
singularity and the existence of the ``singularity model'' given
in mathematics, make the studying of the initial singularity of the
big bang universe feasible. The Ricci flow approach is not only powerful
to study the compact geometry (as Hamilton's and Perelman's seminal
works had shown) but also to the non-compact geometry \citep{1989Ricci,1989Deforming,2005Uniqueness}.
In fact, the classification of a manifold in mathematics is nothing
but equivalent to constructing the complete Hilbert space of the spacetime,
and hence quantizing the spacetime in the language of physics. We
first lay the physics foundation to the Ricci flow of spacetime based
on the notion of quantum reference frame, and we consider the Ricci
flow at least at the Gaussian approximation as a candidate quantum
theory of spacetime. As long as the equivalence principle of Einstein
is consistently generalized to the quantum level, the Ricci flow of
the quantum spacetime can also be a candidate quantum theory for the
gravity. The correctness of the framework depends on its mathematical
consistency and the validity of applying it to explain and predict
observations.

This framework of quantum spacetime is capable to discuss several
fundamental issues of quantum gravity, for instance, the cosmological
constant problem \citep{Luo2014The,Luo2015Dark,Luo:2015pca,Luo:2019iby},
the trace anomaly \citep{Luo:2021zpi}, the thermodynamics of the
quantum spacetime \citep{Luo:2022statistics} and the modified gravity\citep{Luo:2022ywl}.
Thus the main motivation and goal of the paper is trying to apply
the framework to another important touchstone of a quantum gravity:
the initial singularity of the universe and its possible early inflationary
epoch.

The structure of the paper is as follows. In section II, we briefly
review and introduce the background of the Ricci flow of spacetime
based on the notion of quantum reference frame. In section III, we
focus on the early universe within framework of the Ricci flow limit
of the quantum spacetime. The connection between a local singularity
formation of the Ricci flow and the early universe is discussed in
III-A; in III-B, we show that, the vicinity of the early epoch singularity
can be modeled by a solution of a gradient shrinking Ricci soliton
(GSRS) equation, the solution naturally gives rise to an inflationary
universe without any speculative inflaton fields; the slightly deviation
from the exact deSitter inflation metric is discussed in III-C where
the slow roll parameters can be calculated by the Ricci flow; and
how the inflation comes to an end is discussed in III-D; the power
spectrums of the scalar and tensor primordial perturbations are calculated
in III-E and be compared with the textbook standard inflation theory;
in III-F, two important quantities: the manifolds density $u_{*}$
and Hubble rate $H_{*}$ during the early universe in the primordial
perturbation power spectrum is estimated to give more practical predictions
to the power spectrums. Finally, we summarize and conclude the paper
in section IV.

\section{Ricci Flow of Quantum Spacetime}

\subsection{Quantum Reference System}

The quantum reference frame (see e.g. \citep{1984PhRvD..30..368A,rovelli1991quantum,dickson2004view,angelo2011physics,Flaminia2019Quantum}
and references therein) is the conceptual foundation of the theory.
In this theory, an under-study-system that is relative to the quantum
reference system is described by a quantum state $|\psi\rangle$,
and the quantum reference system is also described by a quantum state
$|X\rangle$. Then the whole system is given by an entangled state
\begin{equation}
|\psi[X]\rangle=\sum_{ij}\alpha_{ij}|\psi\rangle_{i}\otimes|X\rangle_{j}\label{eq:entangled state}
\end{equation}
in the Hilbert space $\mathcal{H}_{\psi}\otimes\mathcal{H}_{X}$. 

To explain the framework, let us first take a 1-dimensional quantum
clock-time as a preliminary and simple example of a reference system,
and then the example can be easily generalized to 4-dimensional quantum
spacetime reference frame. In Newtonian mechanics, the under-study-system
are the coordinates $\mathbf{X}$ of a particle, as functions of the
absolute time $t$, i.e. $\mathbf{X}(t)$, called the equation of
motion of the particle. To keep the speed of light a constant in any
reference frame, Einstein pointed out that the coordinates of the
particle in a moving frame are, instead, $\mathbf{X}(\tau),T(\tau)$,
including a physical clock-time $T(\tau)$ in each different moving
frame, where $\tau$ is certain global parameter (e.g. proper time).
An observer sees the equations of motion of the particle being reference
to the relative time $T(\tau)$ is then given by functional $\mathbf{X}[T(\tau)]$,
which generalizes the functions $\mathbf{X}(t)$ in Newtonian mechanics.
The functionals describe the relation between the coordinate $\mathbf{X}$
and the physical clock-time $T$. For example, the action of the coordinate
$X$ of the particle w.r.t. the proper time $\tau$ (a global parameter)
is given by a standard particle action with mass $m_{\mathbf{X}}$
and potential $V(\mathbf{X})$
\begin{equation}
S_{\mathbf{X}}=\int d\tau\left[\frac{1}{2}m_{\mathbf{X}}\left(\frac{d\mathbf{X}}{d\tau}\right)^{2}-V(\mathbf{X})\right].\label{eq:action of X}
\end{equation}
While time is imagined as an ideal and fiducial motion that other
more complex motions are relative to. So we could also consider the
clock-time $T$ as the coordinate of a pointer of a physical clock,
which is considered uniformly moving and hence free. Thus the action
of the coordinate $T$ of the pointer w.r.t. the proper time $\tau$
(global parameter) is given by a standard free particle action without
potential
\begin{equation}
S_{T}=\int d\tau\left[\frac{1}{2}m_{T}\left(\frac{dT}{d\tau}\right)^{2}\right]\label{eq:action of clock}
\end{equation}
where $m_{T}$ is the mass of the clock pointer. The global parameter
$\tau$ can be interpreted as the proper time of the lab. Since the
under-study particle has no interaction with the clock after initial
calibration between them, so the action of the whole system is a sum
of them without their interaction,
\begin{equation}
S[\mathbf{X},T]=S_{\mathbf{X}}+S_{T}.\label{eq:action X+T}
\end{equation}

At the quantum level, the description of the particle is replaced
by a quantum state $|\mathbf{X}(\tau)\rangle$ given by the Hamiltonian
of the action $S_{\mathbf{X}}$, and the clock is also described by
a quantum state $|T\rangle$ from the Hamiltonian of the action $S_{T}$.
Then the evolution functional $\mathbf{X}[T(\tau)]$ in classical
physics, which describes a correspondence between functions $T(\tau)$
and $\mathbf{X}(\tau)$, is now replaced by a quantum version of correspondence
between the states $|T(\tau)\rangle$ and $|\mathbf{X}(\tau)\rangle$,
i.e. an entangled state 
\begin{equation}
|\mathbf{X}[T]\rangle=\sum_{\tau}\alpha_{\tau}|\mathbf{X}(\tau)\rangle\otimes|T(\tau)\rangle
\end{equation}
which is a simple example of (\ref{eq:entangled state}). The state
predicts the output of the joint measurements of the clock time $|T(\tau)\rangle$
and the coordinate $|\mathbf{X}(\tau)\rangle$ of the particle at
the same ``time'' $\tau$, in this sense the equation of motion
of the particle $\mathbf{X}[T(\tau)]$ is measured at the quantum
level. Different from that classical physics predicts a deterministic
relation $\mathbf{X}[T(\tau)]$, the quantum mechanics predicts a
probabilistic state $|\mathbf{X}[T]\rangle=\sum_{\tau}\alpha_{\tau}|\mathbf{X}(\tau)\rangle\otimes|T(\tau)\rangle$.
Following the standard Copenhagen interpretation of the quantum state,
$|\alpha_{\tau}|^{2}$ is a joint probability of the particle state
$|\mathbf{X}(\tau)\rangle$ and the clock state $|T(\tau)\rangle$
happening at the same ``time'' $\tau$. The (conditional) probability
of the particle at $|\mathbf{X}(\tau)\rangle$ in the condition when
the clock is at the state $|T(\tau)\rangle$ can also be given by
$|\alpha_{\tau}|^{2}/|c_{\tau}|^{2}$, where $|c_{\tau}|^{2}$ is
the individual probability of the clock at the state $|T(\tau)\rangle$.
And because the entangled state is inseparable, the joint probability
$|\alpha_{\tau}|^{2}$ will not be a direct product of the individual
probability values of $|\mathbf{X}(\tau)\rangle$ and $|T(\tau)\rangle$,
so the conditional probability $|\alpha_{\tau}|^{2}/|c_{\tau}|^{2}$
will not be an individual probability of the particle state $|\mathbf{X}(\tau)\rangle$.
In this sense, the entangled state (\ref{eq:entangled state}) describes
a ``relational state'' between the under-study-system (e.g. the
coordinate of the particle) and the quantum reference system (e.g.
the clock time), rather than an ``absolute state'' in textbook quantum
mechanics. 

In the semi-classical approximation, the quantum fluctuation $\langle T^{2}\rangle-\langle T\rangle^{2}=\langle\delta T^{2}\rangle$
is ignored, so that the clock-time $T$ can be seen as a c-number
parameter $\langle T\rangle$, i.e. a delta wavefunction peaked at
$\langle T\rangle$, then the action is rewritten as
\begin{align}
S[\mathbf{X},T]\overset{(1)}{\approx}S[\mathbf{X}\left(\langle T\rangle\right)] & =\int dT\left\Vert \frac{d\tau}{dT}\right\Vert \left\{ \left(\frac{dT}{d\tau}\right)^{2}\left[\frac{1}{2}m_{\mathbf{X}}\left(\frac{\delta\mathbf{X}}{\delta T}\right)^{2}+\frac{1}{2}m_{T}\right]-V(\mathbf{X})\right\} \nonumber \\
 & =\int dT\left[\frac{1}{2}M_{\mathbf{X}}\left(\frac{\delta\mathbf{X}}{\delta T}\right)^{2}-V(\mathbf{X})+\textrm{const}\right]\label{eq:semi-classical action of X}
\end{align}
where $\overset{(1)}{\approx}$ means the approximation is at the
1st order/semi-classical approximation compared with the 2nd order/Gaussian
approximation in the coming discussions, $\left\Vert \frac{d\tau}{dT}\right\Vert $
is a Jacobian determinant, and $M_{\mathbf{X}}=m_{\mathbf{X}}\left\Vert \frac{d\tau}{dT}\right\Vert \left(\frac{dT}{d\tau}\right)^{2}=m_{\mathbf{X}}\frac{d\langle T\rangle}{d\tau}$
is the effective mass of the particle. When the semi-classical clock
time runs at exactly the same rate with the proper time $\tau$, then
$M_{\mathbf{X}}=m_{\mathbf{X}}$, and hence the semi-classical action
recovers the action (\ref{eq:action of X}) up to a constant, only
the lab's parameter time $\tau$ is replaced by the mean value of
the physical clock time $\langle T\rangle$, and the derivative $\frac{d}{d\tau}$
is replaced by the functional derivative $\frac{\delta}{\delta\langle T\rangle}$,
the entangled state $|\mathbf{X}[T]\rangle$ recovers the textbook
quantum absolute state $|\mathbf{X}(\tau)\rangle$.

An important observation is that $T$ interpreted as the clock time
is quadratic in the action (\ref{eq:action X+T}), so its first functional
derivative of $T$ vanishes, $\frac{\delta S[\mathbf{X},T]}{\delta T}=\langle E\rangle=0$,
and hence the Hamiltonian corresponding to the action is zero, that
is to say that the Schrodinger equation of the whole system is in
fact a timeless wave equation with zero Hamiltonian (some literature
call it Wheeler-DeWitt equation). There is no external time here,
on the contrary, the equation is used to defined time $T$ which monitors
the under-study particle system $\mathbf{X}$. However, the second
derivative of $T$ is not zero, so when the 2nd order fluctuation
$\langle\delta T^{2}\rangle\neq0$ of the clock time is taken into
account, the energy fluctuation $\sqrt{\langle\delta E^{2}\rangle}\neq0$
gives rise to a correct order of vacuum energy driving a late epoch
acceleration expansion of the universe \citep{Luo2014The}. 

\subsection{Non-Linear Sigma Model (NLSM) for the Quantum Spacetime Frame Fields}

In the subsection, we generalize the quantum clock time to the quantum
spacetime reference frame. To locate the coordinates of an event,
the number of references must be generalized from 1 (one clock) to
at least $D=4$ (3 rods plus one clock), i.e. $X_{\mu}=(X_{0},X_{1},X_{2},X_{3})$.
To contain these 4 frame fields in a lab, the number of the global
parameter must be generalized from one (lab's proper time $\tau$)
to $d=4-\epsilon$ (lab's Minkowskian/Euclidean background), i.e.
$x=(x_{0},x_{1},x_{2},x_{3})$. From the mathematical point of view,
the D-dimensional frame fields is a manifold $X_{\mu}$, which is
a non-linear differentiable mapping $X(x)$ from a local coordinate
patch $x\in\mathbb{R}^{d}$ to a D-manifolds $X\in M^{D}$. For the
same logic of the previous quantum reference system, we can also consider
the frame fields $X$ as the under-study-system w.r.t. the lab's coordinates
$x$ as the reference system, then the local mapping (i.e frame field)
$X(x)$ at the quantum level is also given by an entangled state $\sum_{x}\alpha_{x}|X\rangle\otimes|x\rangle$,
rather than a direct product state, describing their local relation. 

At the moment, the entanglement between the under-study-system (frame
fields $X$) and a local reference system (local lab frame $x$) has
not directly related to the inflation (the main subject of the paper),
the inflation comes from the singularity developed by the local mapping
and the entanglement in a indirect way, next we will introduce the
dynamical theory and corresponding RG-flow of such local mapping and
the entanglement. 

The mapping in physics is usually realized by a kind of fields theory,
the non-linear sigma model (NLSM) \citep{gell1960axial,friedan1980nonlinear,Friedan1980,codello2009fixed,2015Non}
\begin{equation}
S[X(x)]=\frac{1}{2}\lambda\int d^{d}xg_{\mu\nu}\frac{\partial X^{\mu}}{\partial x_{a}}\frac{\partial X^{\nu}}{\partial x_{a}},\label{eq:NLSM}
\end{equation}
which is a generalization of the free clock time action (\ref{eq:action of clock}).
$|X\rangle$ in (\ref{eq:entangled state}) is the eigenstate of the
Hamiltonian of this action. $\lambda$ is the only input constant
with dimension of energy density $[L^{-d}]$ taking the value (\ref{eq:critical density}),
which is the generalization of the clock mass. $x_{a}$ called the
base space in NLSM, representing the lab's wall and clock frame as
the starting reference, which are the generalization of the single
proper time $\tau$ in (\ref{eq:action of clock}). In our common
sense, the lab's spacetime frame has dimension $d=4-\epsilon$, and
is considered fiducial, classical, flat and external with infinite
precision. Note that the NLSM is background signature independent,
so without loss of generality, we consider the base space as an Euclidean
one, i.e. $x\in\mathbb{R}^{d}$ which is better defined when we use
the functional method to quantize the theory in latter section.

The differential mapping or the frame fields $X_{\mu}(x)$ with dimensional
length $[L]$, now is the physical coordinates of a Riemannian or
Lorentzian spacetime $M^{D}$ with generally curved metric $g_{\mu\nu}$,
called the target space in NLSM. When quantizing the theory, we will
promote the frame fields $X_{\mu}$ to D quantum fields, so in the
language of quantum fields theory, $X_{\mu}(x)$ or their duals $X^{\mu}(x)=g^{\mu\nu}X_{\nu}(x)$
are the real defined scalar frame fields.

The under-study-system, without loss of generality, can be generalized
from $\mathbf{X}(\tau)$ (in previous subsection) to a standard scalar
field $\psi(x)$, which shares the lab's background $x$ with the
frame fields $X(x)$. The action is generalized from (\ref{eq:action of X})
to
\begin{equation}
S[\psi(x)]=\int d^{d}x\left[\frac{1}{2}\frac{\partial\psi}{\partial x_{a}}\frac{\partial\psi}{\partial x_{a}}-V(\psi)\right]\label{eq:action of psi}
\end{equation}
where $V(\psi)$ is some potential of the scalar field. $|\psi\rangle$
in (\ref{eq:entangled state}) is the eigenstate of the Hamiltonian
of the action.

Then the total action of the scalar field and the frame fields is
a sum of each system without interaction between them
\begin{equation}
S[\psi,X]=\int d^{d}x\left[\frac{1}{2}\frac{\partial\psi}{\partial x_{a}}\frac{\partial\psi}{\partial x_{a}}-V(\psi)+\frac{1}{2}\lambda g_{\mu\nu}\frac{\partial X^{\mu}}{\partial x_{a}}\frac{\partial X^{\nu}}{\partial x_{a}}\right],
\end{equation}
as the generalization of (\ref{eq:action X+T} or \ref{eq:semi-classical action of X}).
The state of the whole state now gives the entangled state (\ref{eq:entangled state}). 

Because both $\psi$ field and the frame fields $X$ share the common
lab's background spacetime $x$, here they are described w.r.t. the
lab's background spacetime as the starting reference. If $\psi$ field
is considered w.r.t. the physical frame fields $X$, the action can
be rewritten by transforming from $x$ to $X$. At the semi-classical
level, when the fluctuation of the frame fields $\langle\delta X^{2}\rangle$
can be ignored, it is simply a coordinates transformation $x\rightarrow\langle X\rangle$,
\begin{align}
S[\psi,X]\overset{(1)}{\approx}S[\psi(\langle X\rangle)] & =\int d^{D}X\sqrt{|\det g^{(1)}|}\left[\frac{1}{4}\left\langle g_{\mu\nu}^{(1)}\frac{\partial X^{\mu}}{\partial x_{a}}\frac{\partial X^{\nu}}{\partial x_{a}}\right\rangle \left(\frac{1}{2}g^{(1)\mu\nu}\frac{\delta\psi}{\delta X^{\mu}}\frac{\delta\psi}{\delta X^{\nu}}+2\lambda\right)-V(\psi)\right]\nonumber \\
 & =\int d^{D}X\sqrt{|\det g^{(1)}|}\left[\frac{1}{2}g^{(1)\mu\nu}\frac{\delta\psi}{\delta X^{\mu}}\frac{\delta\psi}{\delta X^{\nu}}-V(\psi)+2\lambda\right],\label{eq:eff-(1)}
\end{align}
in which $\frac{1}{4}\left\langle g_{\mu\nu}^{(1)}\frac{\partial X^{\mu}}{\partial x_{a}}\frac{\partial X^{\nu}}{\partial x_{a}}\right\rangle =\frac{1}{4}\left\langle g_{\mu\nu}^{(1)}g^{(1)\mu\nu}\right\rangle =\frac{1}{4}D=1$
has been used. The semi-classical action is a generalization of (\ref{eq:semi-classical action of X}).
It is easy to see, under the semi-classical treatment of frame fields
$X$, the classical coordinates transformation reproduces the scalar
field action (\ref{eq:action of psi}) in general spacetime reference
frame coordinates $X$ up to a constant $2\lambda$, lab's parameter
background $x$ is replaced by $\langle X\rangle$, and the derivative
$\frac{\partial}{\partial x_{a}}$ is replaced by the functional derivative
$\frac{\delta}{\delta X^{\mu}}$. $\sqrt{|\det g^{(1)}|}=\left\Vert \frac{dx}{dX}\right\Vert $
is the Jacobian determinant of the coordinate transformation, which
is the generalization of $\left\Vert \frac{d\tau}{dT}\right\Vert $
in previous subsection. Note that the coordinates transformation matrix
must be a square matrix, so at semi-classical level $d$ should be
close to $D=4$, which is seem obviously that the dimension of the
lab is 4. However, it is for quantum and topological reasons that
$d$ must not be exactly 4 but rather $d=4-\epsilon$. In fact, the
value of $d$ is very crucial for the renormalizability of the theory
at the quantum level. It is well known that $d=2$ the NLSM is power
counting and perturbative renormalizable. Although $2<d<4$ is not
power counting and perturbative renormalizable, it is evidence that
the theory is non-perturbative renormalizable. From the topological
point of view, NLSM is a fields model of mapping from the base space
$x$ to the target space $X$, whether NLSM is well-defined at the
quantum or renormalization level depends on whether the mapping is
free from intrinsic topological singularity. For simplicity, we consider
the target space topologically a 4-sphere (after properly Wick rotated),
$M^{D}=S^{4}$, then the homotopy group of the mapping $X:\mathbb{R}^{d}\rightarrow S^{4}$
is $\pi_{d}(S^{4})$. The homotopy group is trivial when $d<4$, i.e.
$\pi_{d<4}(S^{4})=0$, which means that all possible (in the path
integral sense) differentiable mapping $X(x)$ will not meet intrinsic
topological obstacles and hence the mapping is always well-defined
and free from intrinsic singularities. That is the reason we choose
$d=4-\epsilon$ as the dimension of the base space, where $\epsilon$
can be considered as a small regularization parameter to avoid mathematical
singularity at the quantum level. In fact, $d$ as an input parameter
is not an observable of the theory, at the quantum level, $d$ can
even be a fractal dimension due to the ``dimension anomaly''. While
at the classical or semi-classical level, it is no problem if one
roughly considers $d=4$.

When the location of an event is at a long distance scale far beyond
the lab's size, for instance, to the galaxy or cosmic scale, when
the frame fields signal travels along such a long distance and be
read by an observer, the broadening or the variance (2nd order moment)
fluctuation of the frame fields $\langle\delta X^{2}\rangle$ become
unignorable. More precisely, the variance $\langle\delta X^{2}\rangle$
inevitably modifies the quadratic form of distance of the Riemannian/Lorentzian
spacetime 
\begin{equation}
\left\langle \left(\varDelta X\right)^{2}\right\rangle =\langle\varDelta X\rangle^{2}+\langle\delta X^{2}\rangle.\label{eq:dx^2}
\end{equation}
Since a local distance element is usual attributed to the local metric
tensor at the point, so it is also convenient to think of the location
point $X$ being fixed, and the effect of the variance affects only
the metric tensor $g_{\mu\nu}$ at the point. As a consequence, the
quantum expectation value of a metric tensor $g_{\mu\nu}$ is modified
by the 2nd moment quantum fluctuation of the frame fields
\begin{equation}
\langle g^{\mu\nu}\rangle=\left\langle \frac{\partial X^{\mu}}{\partial x_{a}}\frac{\partial X^{\nu}}{\partial x_{a}}\right\rangle =\left\langle \frac{\partial X^{\mu}}{\partial x_{a}}\right\rangle \left\langle \frac{\partial X^{\nu}}{\partial x_{a}}\right\rangle +\frac{1}{2}\frac{\partial^{2}}{\partial x_{a}^{2}}\left\langle \delta X^{\mu}\delta X^{\nu}\right\rangle =g_{(1)}^{\mu\nu}(X)+\delta g_{(2)}^{\mu\nu}(X),\label{eq:g=00003Dg(1)+dg(2)}
\end{equation}
where
\begin{equation}
g_{(1)}^{\mu\nu}(X)=\left\langle \frac{\partial X^{\mu}}{\partial x_{a}}\right\rangle \left\langle \frac{\partial X^{\nu}}{\partial x_{a}}\right\rangle =\langle e_{a}^{\mu}\rangle\langle e_{a}^{\nu}\rangle
\end{equation}
is the 1st order moment (mean value) contribution, $e_{a}^{\mu}$
is the vierbein. For correction from the 2nd order moment or variance
$\delta g_{(2)}^{\mu\nu}$ deforms the metric and the geometry of
the physical spacetime, especially at long distance or cosmic scale.
It is a renormalization or coarse-graining process of the quantum
spacetime. 

\subsection{Metric deformation from Quantum Gaussian Fluctuations: Ricci Flow}

To investigate the quantum correction $\delta g_{(2)}^{\mu\nu}(X)$,
we need to quantize the frame fields at the 2nd order moment or Gaussian
approximation level, if the higher order moments (non-Gaussian fluctuations)
are less important compared to the Gaussian fluctuation. When $\delta g_{(2)}^{\mu\nu}$
is relatively smaller than $g_{(1)}^{\mu\nu}$, it can be given by
a perturbative one-loop calculation \citep{codello2009fixed} of the
NLSM
\begin{equation}
\delta g_{(2)}^{\mu\nu}(X)=\frac{1}{2}\frac{\partial^{2}}{\partial x_{a}^{2}}\langle\delta X^{\mu}\delta X^{\nu}\rangle=-\frac{R_{(1)}^{\mu\nu}(X)}{32\pi^{2}\lambda}\delta k^{2},\label{eq:dg(2)}
\end{equation}
where $R_{(1)}^{\mu\nu}$ is the Ricci curvature given by the 1st
order metric $g_{(1)}^{\mu\nu}$, $k^{2}$ is the cutoff energy scale
of the Fourier components of the frame fields $X_{\mu}$. The validity
of the perturbation calculation $R^{(1)}\delta k^{2}\ll\lambda$ is
actually the validity of the Gaussian approximation. In fact, to recover
the standard General Relativity, $\lambda$ must take the value of
the critical density $\rho_{c}$ of the universe (shown in (\ref{eq:critical density})),
i.e. $\lambda\sim O(H_{0}^{2}/G)$, $H_{0}$ the current Hubble's
constant, $G$ the Newton's constant. For the case when the curvature
is of order of $H_{0}$, the condition $R^{(1)}\delta k^{2}\ll\lambda$
is equivalent to $\delta k^{2}\ll1/G$ which is reliable. For the
case when the curvature is large near a local singularity of a manifold,
in principle, contributions from non-Gaussian fluctuations depicted
by higher powers of the curvature $(R_{\mu\nu}/\lambda)^{n>1}$ may
become important, the Gaussian approximation may be fail. In fact,
depending on what is our interested physics and how close the period
producing our interested physics is to the local singularity, the
validity of the Gaussian approximation is a subtle issue in the early
epoch of the universe which will be discussed in the next section.
It is also worth stressing that here the metric fluctuation $\delta g_{(2)}^{\mu\nu}$
is not directly related to the observed tensor modes of metric perturbation
in inflation. It will give a RG-flow to the spacetime, and finally
gives rise to the curvature pinching near the singularity of the early
universe (see section III).

The equation (\ref{eq:dg(2)}) is actually a RG equation of the target
space, i.e. the physical spacetime, in mathematics it called the Ricci
flow equation (some reviews see e.g. \citep{chow2004ricci,chow2006hamilton,topping2006lectures})
\begin{equation}
\frac{\partial g^{\mu\nu}}{\partial t}=2R^{\mu\nu}\quad\textrm{or}\quad\frac{\partial g_{\mu\nu}}{\partial t}=-2R_{\mu\nu}.\label{eq:ricci flow}
\end{equation}
For the same convention in mathematics literature, we often use the
latter to describe a continuous deformation of the tangent spacetime
metric driven by its Ricci curvature. The flow-time interval $\delta t=-\frac{1}{64\pi^{2}\lambda}\delta k^{d-2}$
has dimension of length square $[L^{2}]$ for any $d$, not restricted
to Friedan's original consideration of $d=2+\epsilon$.

For the Ricci curvature is non-linear in metric, the Ricci flow equation
is in analogy to a non-linear ``heat equation'' for the metric,
and flow along $t$ introduces a renormalizing or coarse-graining
process to a spacetime and gravitational system which is highly non-trivial
\citep{carfora1995renormalization,piotrkowska1995averaging,carfora2008ricci,Zalaletdinov:2008ts,Paranjape:2009zu}.
If it is free from local singularities during the flow, there exists
a long flow-time solution in $t\in(-\infty,0)$, which is often called
an ancient solution in mathematics. In this situation, the range of
the t-parameter corresponds to $k\in(0,\infty)$, from $t=-\infty$,
i.e. a short distance (high energy) UV scale $k=\infty$ forwardly
to $t=0$ i.e. a long distance (low energy) IR scale $k=0$. The coarse-grained
metric at scale $t$ is given by being averaged out the fine-grain
or short distance fine structure of the metric. So along t, the manifolds
loss its fine-grain information, so that the flow is irreversible,
that is it has no backwards solution. It is the underlying reason
for the existence of an entropy of a spacetime discussed in \citep{Luo:2022statistics}
and in later subsections. 

Note that in (\ref{eq:dx^2}), (\ref{eq:g=00003Dg(1)+dg(2)}), the
variance of the metric modifies the local quadratic form of spacetime
distance, thus the flow is essentially non-isometry. It is the underlying
reason for the diffeomorphism anomaly of spacetime. The non-isometry
does not affect its topology, so the flow preserves the topology of
the spacetime. However, its local metric, shape and size (volume)
changes during the flow. And there also exists a very special solution
of the Ricci flow called the Ricci Soliton \citep{1988The} in mathematics,
which is a generalization of the notion of quasi-Einstein metric in
physics \citep{Friedan1980}. The solution only deforms its local
volume while keeps its local shape, the solution is self-similar.
The Ricci Soliton, and its more general version, the Gradient Ricci
Soliton, plays the role of the flow limit, are the generalization
of the notion of fixed point in the RG flow and the deSitter metric.
In the Gradient Ricci Soliton, the Gradient Shrinking Ricci Soliton
(GSRS) is a particularly important model in understanding the gravity
at cosmic scale and early epoch, we will see in the next section.

\subsection{Density Matrix of the Frame Fields: induced Ricci-DeTurck Flow}

Because the Ricci flow is a coarse-graining process of the spacetime,
there exist irreversible entropy and diffeomorphism anomaly in the
framework (see next subsection), so density matrix, rather pure state,
is a more proper fundamental notion. By using the density matrix $u$,
the previous (2nd order) results e.g. (\ref{eq:g=00003Dg(1)+dg(2)}),
(\ref{eq:dg(2)}) and hence the Ricci flow (\ref{eq:ricci flow})
can also be given by the expectation value 
\begin{equation}
\langle O\rangle=\langle X|O|X\rangle=\lambda\int d^{4}X\Psi^{*}(X)O\Psi(X)=\lambda\int d^{4}Xu(X)O
\end{equation}
via writing down the wavefunction $\Psi(X)$ or density matrix $u$
of the frame fields explicitly at the Gaussian approximation, where
\begin{equation}
d^{4}X=\sqrt{|\det g_{\mu\nu}|}dX^{0}dX^{1}dX^{2}dX^{3}\label{eq:volume element}
\end{equation}
Remind that at the semi-classical approximation, the frame fields
$X$ is a delta density peaking at its mean value. Thus at the Gaussian
approximation level, finite Gaussian width/2nd moment fluctuation
of $X$ must be introduced as the next order correction to the delta
density. So at the Gaussian approximation, the fundamental solution
of the wave function takes the Gaussian form
\begin{equation}
\Psi[X^{\mu}(x)]=\frac{\left|\det\sigma_{\mu\nu}\right|^{1/4}}{\sqrt{\lambda}(2\pi)^{D/4}|g|^{1/4}}\exp\left[-\frac{1}{4}\left|\left(X^{\mu}(x)-x^{\mu}\right)\sigma_{\mu\nu}\left(X^{\nu}(x)-x^{\nu}\right)\right|\right],
\end{equation}
where the covariant matrix $\sigma_{\mu\nu}(x)$ measures the 2nd
order moment fluctuations of the frame fields at the peaking point
$x=\langle X\rangle$
\begin{equation}
\sigma_{\mu\nu}(x)=\frac{1}{\sigma^{\mu\nu}(x)}=\frac{1}{\left\langle \delta X^{\mu}(x)\delta X^{\nu}(x)\right\rangle }.
\end{equation}
The absolute symbol in the exponential of the wavefunction is for
keeping the Gaussian integral over $X$ positive as a probability,
even in the Lorentzian (signature) spacetime.

The fundamental solution of the wavefunction gives rise to a dimensionless
probability density matrix 
\begin{equation}
u[X^{\mu}(x)]=\Psi^{*}(X)\Psi(X)=\frac{1}{\lambda(2\pi)^{D/2}}\frac{\left|\det\sigma_{\mu\nu}\right|^{1/2}}{\left|\det g_{\mu\nu}\right|^{1/2}}\exp\left[-\frac{1}{2}\left|\left(X^{\mu}(x)-x^{\mu}\right)\sigma_{\mu\nu}\left(X^{\nu}(x)-x^{\nu}\right)\right|\right],\label{eq:u}
\end{equation}
in which $\frac{\sqrt{\left|\det\sigma_{\mu\nu}\right|}}{\lambda(2\pi)^{D/2}\sqrt{|\det g_{\mu\nu}|}}$
is given by the normalization condition 
\begin{equation}
\lambda\int d^{D}X\Psi^{*}(X)\Psi(X)=\lambda\int d^{D}Xu(X)=1.\label{eq:u-normalization}
\end{equation}

There exists an arbitrariness in the density $u(X)$ for different
choices of a diffeomorphism/gauge. Under a diffeomorphism $g_{\mu\nu}\rightarrow\hat{g}_{\mu\nu}$,
$u(X)$ is transformed corresponding to a diffeomorphism of the covariant
matrix
\begin{equation}
\sigma_{\mu\nu}\rightarrow\hat{\sigma}_{\mu\nu}=\sigma_{\mu\nu}+\nabla_{\mu}\nabla_{\nu}h.\label{eq:2nd moment}
\end{equation}
where $h$ is certain transformation function.

Following the statistical interpretation of wavefunction with the
normalization condition (\ref{eq:u-normalization}), density matrix
$u(X^{0},X^{1},X^{2},X^{3})$ measures the probability density of
finding the frame fields particles in the volume $d^{D}X$. During
the Ricci flow along $t$, the volume $\Delta V_{t}$ in which the
density is averaged also flows, so the density is coarse-grained in
the volume $\Delta V_{t}$ at the scale $t$. If we consider the volume
of the lab is rigid and fixed by $\lambda\int d^{4}x=1$, so
\begin{equation}
u[X(x),t]=\frac{d^{4}x}{d^{D}X_{t}}=\lim_{\Delta V_{t}\rightarrow0}\frac{1}{\Delta V_{t}}\int_{\Delta V}1\cdot d^{4}x=\langle1\rangle_{\Delta V_{t}\rightarrow0}.\label{eq:coarse-grain density}
\end{equation}
Thus the density $u(X,t)$ can be interpreted as a coarse-grained
density at the scale $t$ w.r.t. a fine-grained unit density in the
lab at UV $t\rightarrow-\infty$.

The coarse-grained density $u(X,t)$ not only has statistic meaning,
playing a central role in analyzing the statistic physics \citep{Luo:2022statistics}
of the frame fields, but also has profound geometric meaning, generalizing
the Riemannian/Lorentzian manifolds $(M^{D},g)$ to a density manifolds
$(M^{D},g,u)$ \citep{Morgan2009Manifolds,2016arXiv160208000W,Corwin2017Differential},
in which $u$ also called a manifold density in mathematics. 

$u(X,t)$ associates a manifold density to each point $\langle X\rangle$
in a manifold. And it is worth stressing that $u$ is not equivalent
to scaling the metric conformally by a factor, because in this case
the integral measure of 4-volume or 3-volume in the expectation $\langle O\rangle=\lambda\int d^{D}XuO$
would scale by different powers. Since Gaussian $u$ density fuzzes
the coordinates of the manifolds and hence deforms the metric and
curvature in the density manifolds. Thus an important observation
is that $u$ density generalizes the notion of curvature in the density
manifolds. Indeed, there are various useful generalizations (e.g.
see \citep{Akbar_2009}) of the Ricci curvature to the density manifolds,
a widely accepted version is the Bakry-Emery generalization \citep{1985Diffusions}
\begin{equation}
R_{\mu\nu}\rightarrow R_{\mu\nu}-\nabla_{\mu}\nabla_{\nu}\log u,\label{eq:R-ddlogu}
\end{equation}
which is also used in Perelman's seminal papers. Such generalization
has many advantages in physics, for instance, the 2nd moment fluctuation
eq.(\ref{eq:2nd moment}) of spacetime encoded in the $u$ density
has more direct relation to curvature and gravity at the quantum level.
We will use the definition throughout the paper. By using the generalized
Ricci curvature, the Ricci flow for Riemannian manifolds $(M^{D},g)$
is generalized to the Ricci-DeTurck flow \citep{deturck1983deforming}
for the density manifolds $(M^{D},g,u)$
\begin{equation}
\frac{\partial g_{\mu\nu}}{\partial t}=-2\left(R_{\mu\nu}-\nabla_{\mu}\nabla_{\nu}\log u\right),\label{eq:ricci-deturk}
\end{equation}
which is nothing but actually equivalent to the Ricci flow equation
(\ref{eq:ricci flow}) up to a diffeomorphism. Different from the
standard Ricci flow, the Ricci-DeTurck flow has the advantage that
it turns out to be a gradient flow of some monotonic functionals introduced
by Perelman, so the Ricci-DeTurck flow equation, instead of the Ricci
flow equation, is a fundamental equation to the physical spacetime
with density at the quantum level. 

Remind that the eq.(\ref{eq:coarse-grain density}) also gives a volume
constraint to the fiducial spacetime (the lab), the density $u(X,t)$
in this sense cancels the flow of $\sqrt{|\det g_{\mu\nu}|}$, so
we have
\begin{align}
\frac{\partial}{\partial t}\left(u\sqrt{|\det g_{\mu\nu}|}\right) & =0.\label{eq:volume constraint}
\end{align}
The relation, together with the Ricci-DeTurck flow equation (\ref{eq:ricci-deturk}),
directly gives rise to the flow equation of the density
\begin{equation}
\frac{\partial u}{\partial t}=\left(R-\Delta\right)u,
\end{equation}
in which $\Delta$ is the Laplacian of the 4-spacetime. And this equation
is in analogy to the irreversible Boltzmann's equation for his distribution
function of dilute gas. However, note the minus sign in front of the
Laplacian operator, it is a backwards heat-like equation of $u$.
It seems that the solution of the backwards heat flow does not exist.
But we also note that if one flows the manifolds to certain IR scale
$t_{s}$, and at the scale $t_{s}$ one can certainly choose an appropriate
$u(t_{s})=u_{0}$ arbitrarily (up to a diffeomorphism gauge) and flows
it backwards in $\tau=t_{s}-t$, and hence one obtains a solution
$u(\tau)$ of the backwards equation. In the situation that if the
flow is considered free from global singularities for the trivialness
of the homotopy group $\pi_{d<4}(S^{4})=0$, we simply consider $t_{s}=0$,
so
\begin{equation}
\tau=-t=\frac{1}{64\pi^{2}\lambda}k^{2}\in(0,\infty).\label{eq:tau}
\end{equation}
As a consequence, the density satisfies a heat-like equation in terms
of the backwards flow time $\tau$
\begin{equation}
\frac{\partial u}{\partial\tau}=\left(\Delta-R\right)u,\label{eq:u-equation}
\end{equation}
which indeed admit a solution along $\tau$, it is often called the
conjugate heat equation in mathematics. 

At this point, (\ref{eq:u-equation}) together with (\ref{eq:ricci-deturk}),
the mathematical problem of the Ricci flow of a Riemannian/Lorentzian
manifolds is generalized to a coupled equations
\begin{equation}
\begin{cases}
\frac{\partial g_{\mu\nu}}{\partial t}=-2\left(R_{\mu\nu}-\nabla_{\mu}\nabla_{\nu}\log u\right)\\
\frac{\partial u}{\partial\tau}=\left(\Delta-R\right)u\\
\frac{d\tau}{dt}=-1
\end{cases}
\end{equation}
and hence the pure Riemannian/Lorentzian manifolds $(M^{D},g)$ is
generalized to a density manifolds $(M^{D},g,u)$ with the constraint
(\ref{eq:u-normalization}).

\subsection{Diffeomorphism Anomaly}

As is shown in previous subsection, an important feature of the Ricci
flow or the Ricci-DeTurck flow to the manifolds is that the quantum
fluctuation does not preserve the quadratic form of the distance of
the manifolds. The non-isometry feature induces a breakdown of diffeomorphism
or general coordinate transformation at the quantum level, namely
the diffeomorphism anomaly \citep{Luo:2021zpi}. 

We consider the functional quantization of the pure frame fields,
the partition function is
\begin{equation}
Z(M^{D})=\int[\mathcal{D}X]\exp\left(-S[X]\right)=\int[\mathcal{D}X]\exp\left(-\frac{1}{2}\lambda\int d^{4}xg^{\mu\nu}\partial_{a}X_{\mu}\partial_{a}X_{\nu}\right),
\end{equation}
in which,without loss of generality, the base spacetime is taken as
Euclidean for it is better defined in path integral, and the final
result is independent to the signature, i.e. the same for Minkowskian. 

We have seen that the behavior of the Ricci flow (\ref{eq:ricci flow})
for the tangent and cotangent spacetime is opposite, for the same
convention in mathematics, here we consider the general coordinate
transformation of the tangent spacetime,
\begin{equation}
X_{\mu}\rightarrow\hat{X}_{\mu}=\frac{\partial\hat{X}_{\mu}}{\partial X_{\nu}}X_{\nu}=e_{\mu}^{\nu}X_{\nu}.
\end{equation}
The coordinate transformation does not change the action $S[X]=S[\hat{X}]$,
but the measure of the functional integral changes
\begin{align}
\mathcal{D}\hat{X} & =\prod_{x}\prod_{\mu=0}^{D-1}d\hat{X}_{\mu}(x)=\prod_{x}\epsilon_{\mu\nu\rho\sigma}e_{\mu}^{0}e_{\nu}^{1}e_{\rho}^{2}e_{\sigma}^{3}dX_{0}(x)dX_{1}(x)dX_{2}(x)dX_{3}(x)\nonumber \\
 & =\prod_{x}\left|\det e(x)\right|\prod_{x}\prod_{a=0}^{D-1}dX_{a}(x)=\left(\prod_{x}\left|\det e(x)\right|\right)\mathcal{D}X,
\end{align}
where
\begin{equation}
\epsilon_{\mu\nu\rho\sigma}e_{\mu}^{0}e_{\nu}^{1}e_{\rho}^{2}e_{\sigma}^{3}=\left|\det e_{\mu}^{a}\right|=\sqrt{\left|\det g_{\mu\nu}\right|}
\end{equation}
is the Jacobian of the coordinate transformation. In fact, the Jacobian
is a local relative volume element $d\mu(\hat{X}_{\mu})$ w.r.t. the
fiducial one $d\mu(X_{a})$. Remind that the normalization condition
(\ref{eq:u-normalization}) defines a fiducial volume element $ud^{4}X\equiv ud\mu(\hat{X}_{\mu})$,
thus the Jacobian actually measures the frame fields density matrix
\begin{equation}
u(\hat{X})=\frac{d\mu(X_{a})}{d\mu(\hat{X}_{\mu})}=\left|\det e_{a}^{\mu}\right|=\frac{1}{\left|\det e_{\mu}^{a}\right|}.\label{eq:volume form}
\end{equation}

The absolute value symbol used in the determinant is to keep $u$
and hence the volume element positive defined, even in the Lorentzian
signature spacetime. Otherwise, for the Lorentzian case, there must
introduce extra imaginary factor $i$ into (\ref{eq:parametrize u})
to preserve the normalization condition (\ref{eq:u-normalization}).
It is a natural generalization from a 3-space density of Perelman
to a 4-spacetime version with Lorentzian signature. It is such definition
of the volume form for the Lorentzian 4-spacetime ensures the formalism
of the framework formally identifies with Perelman's standard formalism
for the 3-space in Euclidean signature. From this observation, we
can see that the manifolds density $u$ encodes one of the most important
information of a manifold, i.e. the local volume ratio, including
not only the classical volume ratio (coming from the classical general
coordinates transformation) but also the quantum counterpart (coming
from the Ricci flow).

We could parameterize the solution $u$ in terms of
\begin{equation}
u(\hat{X})=\frac{1}{\lambda(4\pi\tau)^{D/2}}e^{-f(\hat{X},\tau)}.\label{eq:parametrize u}
\end{equation}
Then by using it, the partition function $Z(M^{D})$ under the coordinate
transformation gives
\begin{align}
Z(\hat{M}^{D}) & =\int[\mathcal{D}\hat{X}]\exp\left(-S[\hat{X}]\right)=\int\left(\prod_{x}\left|\det e\right|\right)[\mathcal{D}X]\exp\left(-S[X]\right)\nonumber \\
 & =\int\left(\prod_{x}e^{f+\frac{D}{2}\log(\lambda^{2/D}4\pi\tau)}\right)[\mathcal{D}X]\exp\left(-S[X]\right)\nonumber \\
 & =\exp\left(\lambda\int d^{4}x\left[f+\frac{D}{2}\log(\lambda^{2/D}4\pi\tau)\right]\right)\int[\mathcal{D}X]\exp\left(-S[X]\right)\nonumber \\
 & =\exp\left(\lambda\int_{\hat{M}^{D}}d^{D}Xu\left[f+\frac{D}{2}\log(\lambda^{2/D}4\pi\tau)\right]\right)\int[\mathcal{D}X]\exp\left(-S[X]\right)
\end{align}
Finally, we can see that the change of the partition function, known
as the anomaly, is
\begin{equation}
Z(\hat{M}^{D})=e^{\lambda N(\hat{M}^{D})}Z(M^{D}),\label{eq:Z->Zhat}
\end{equation}
in which $N(\hat{M}^{D})$ is nothing but a Shannon entropy in terms
of the manifolds density $u$
\begin{equation}
N(\hat{M}^{D})=\int_{\hat{M}^{D}}d^{D}Xu\left[f+\frac{D}{2}\log(\lambda^{2/D}4\pi\tau)\right]=-\int_{\hat{M}^{D}}d^{D}Xu\log u.
\end{equation}
Because $u$ in (\ref{eq:volume form}) is real defined, the Shannon
entropy or the anomaly is real, even for the Lorentzian signature.
It is a general result of the Ricci flow of spacetime. 

Without loss of generality, if we simply consider the under-transformed
coordinates $X_{a}$ to be the coordinates of the fiducial lab $x_{a}$,
so they can be treated as classical parameter coordinates. In this
situation the classical action of NLSM is simply $\frac{D}{2}$, a
topological invariant, i.e.
\begin{equation}
\exp\left(-S_{cl}\right)=\exp\left(-\frac{1}{2}\lambda\int d^{4}xg^{\mu\nu}\partial_{a}x_{\mu}\partial_{a}x_{\nu}\right)=\exp\left(-\frac{1}{2}\lambda\int d^{4}xg^{\mu\nu}g_{\mu\nu}\right)=e^{-\frac{D}{2}}.
\end{equation}
Thus the total partition function (\ref{eq:Z->Zhat}) takes a simple
form
\begin{equation}
Z(\hat{M}^{D})=e^{\lambda N(\hat{M}^{D})-\frac{D}{2}}.\label{eq:frame-partition}
\end{equation}

In the Ricci flow limit, i.e. the Gradient Shrinking Ricci Soliton
(GSRS) configuration
\begin{equation}
R_{\mu\nu}+\nabla_{\mu}\nabla_{\nu}f=\frac{1}{2\tau}g_{\mu\nu},\label{eq:shrinker}
\end{equation}
the covariance matrix $\sigma^{\mu\nu}$ as the 2nd order moment of
the frame fields with a IR cutoff $k$ is simply proportional to the
metric
\begin{equation}
\frac{1}{2}\sigma_{\star}^{\mu\nu}=\frac{1}{2}\langle\delta X^{\mu}\delta X^{\nu}\rangle=\frac{1}{2\lambda}g^{\mu\nu}\int_{0}^{|p|=k}\frac{d^{4}p}{(2\pi)^{4}}\frac{1}{p^{2}}=\frac{k^{2}}{64\pi^{2}\lambda}g^{\mu\nu}=\tau g^{\mu\nu},\label{eq:sigma=00003Dtau*g}
\end{equation}
and then
\begin{equation}
\sigma_{\star\mu\nu}=\left(\sigma_{\star}^{\mu\nu}\right)^{-1}=\frac{1}{2\tau}g_{\mu\nu},\label{eq:shrinking term}
\end{equation}
which means a uniform Gaussian broadening is achieved, i.e. its covariant
gradient vanishes $\nabla_{\rho}\sigma_{\star\mu\nu}=0$. The subscript
``$\star$'' represents the Ricci flow limit at which the Shannon
entropy $N$ approaches to its maximum value $N_{\star}$, and the
density matrix
\begin{equation}
u(X)=\frac{\left|\det\sigma_{\mu\nu}\right|^{1/2}}{\lambda(2\pi)^{D/2}|\det g_{\mu\nu}|^{1/2}}\exp\left(-\frac{1}{2}\left|X^{\mu}\sigma_{\mu\nu}X^{\nu}\right|\right),\label{eq:general u}
\end{equation}
becomes a Maxwell-Boltzmann form of density 
\begin{equation}
u_{\star}(X)=\frac{1}{\lambda(4\pi\tau)^{D/2}}\exp\left(-\frac{1}{4\tau}\left|g_{\mu\nu}X^{\mu}X^{\nu}\right|\right)\label{eq:u*}
\end{equation}
in the limit, in analogy to a ``thermoequilibrium'' state of spacetime
\citep{Luo:2022statistics}. We can also define a relative density
$\tilde{u}$ as the general density $u(X)$ w.r.t. the ``thermoequilibrium''
density $u_{\star}(X)$ in the limit
\begin{equation}
\tilde{u}(X)=\frac{u(X)}{u_{\star}(X)}.
\end{equation}
By using the relative density, a relative Shannon entropy $\tilde{N}$
can also be defined by
\begin{equation}
\tilde{N}(M^{D})=-\int d^{D}Xu\log\tilde{u}=-\int d^{D}Xu\log u+\int d^{D}Xu_{\star}\log u_{\star}=N-N_{\star}=-\log Z_{P}\le0,\label{eq:perelman-partition}
\end{equation}
where $Z_{P}$ is Perelman's partition function 
\begin{equation}
\log Z_{P}=\int_{M^{D}}d^{D}Xu\left(\frac{D}{2}-f\right)\ge0,
\end{equation}
and $N_{\star}$ is the maximum Shannon entropy
\begin{equation}
N_{\star}=-\int d^{D}Xu_{\star}\log u_{\star}=\int d^{D}Xu_{\star}\frac{D}{2}\left[1+\log(\lambda^{2/D}4\pi\tau)\right]=\frac{D}{2\lambda}\left[1+\log(\lambda^{2/D}4\pi\tau)\right].
\end{equation}

For the reason that the relative Shannon entropy is real, the change
of the partition function under diffeomorphism is in general non-unitary.
Perelman defined the F-functional 
\begin{equation}
\mathcal{F}=\frac{dN}{d\tau}=-\frac{dN}{dt}=\int_{M^{D}}d^{D}Xu\left(R+\left|\nabla f\right|^{2}\right)
\end{equation}
with its maximum value at GSRS limit
\begin{equation}
\mathcal{F}_{\star}\equiv\mathcal{F}(u_{\star})=\frac{dN_{\star}}{d\tau}=\frac{D}{2\lambda\tau}.\label{eq:F*}
\end{equation}
It is easy to show that $\mathcal{F}$ is monotonic non-decreasing
along t, because
\begin{align}
\frac{d\mathcal{F}}{dt} & =2\int d^{D}Xu\left|R_{\mu\nu}-\nabla_{\mu}\nabla_{\nu}\log u\right|^{2}\ge\frac{2}{D}\int d^{D}Xu\left|R-\varDelta\log u\right|^{2}\nonumber \\
 & \ge\frac{2}{D}\lambda\left[\int d^{D}Xu\left(R-\varDelta\log u\right)\right]^{2}=\frac{2}{D}\lambda\mathcal{F}^{2}\ge0.
\end{align}
Thus the relative Shannon entropy is monotonic non-decreasing along
the Ricci flow (along $t$),
\begin{equation}
\frac{d\tilde{N}(\hat{M}^{D})}{dt}=-\tilde{\mathcal{F}}\ge0,\label{eq:analog H-theorem}
\end{equation}
where $\tilde{\mathcal{F}}=\mathcal{F}-\mathcal{F}_{\star}\le0$ is
the GSRS-normalized F-functional.

\subsection{Effective Action of Gravity and Non-Singular Ricci Flow}

If the equivalence principle is generalized to the quantum level,
the quantum fluctuation of the frame fields should not only be a property
of frame fields themselves but also be interpreted as universal property
of the spacetime. In the interpretation, the mean value of the frame
fields $\langle X\rangle$ measures the classical coordinate of the
spacetime, the quantum fluctuation $\langle\delta X^{2}\rangle$ of
the frame fields measures the variance or fuzziness of the coordinates,
and hence gravity as a relational phenomenon between different frames
emerges from the frame fields system at the quantum level.

By using $\tilde{N}(M^{D})$, the total partition function (\ref{eq:frame-partition})
now is written as 
\begin{equation}
Z(M^{D})=\frac{e^{\lambda N-\frac{D}{2}}}{e^{\lambda N_{\star}}}=e^{\lambda\tilde{N}-\frac{D}{2}}=Z_{P}^{-\lambda}e^{-\frac{D}{2}}=\exp\left[\lambda\int_{M^{D}}d^{D}Xu\left(f-D\right)\right].\label{eq:relative-partition}
\end{equation}

The relative Shannon entropy $\tilde{N}$ playing the role of the
anomaly vanishes at GSRS limit at IR scale, but in general it is non-zero
at lab scale up to UV. Since the lab's volume is considered fiducial
and fixed by $\lambda\int d^{4}x=1$, so the anomaly must be canceled
at the lab scale up to UV. The physical requirement leads to the counter
term $\nu(M_{\tau=\infty}^{D})$ which finally appears as the cosmological
constant. The monotonicity of $\tilde{N}$ implies
\begin{equation}
\nu(M_{\tau=\infty}^{D})=\lim_{\tau\rightarrow\infty}\lambda\tilde{N}(M^{D},u,\tau)=\lambda\left[\tilde{N}_{UV}(\hat{M}_{UV}^{D})-\tilde{N}_{IR}(\hat{M}_{IR}^{D})\right]<0.
\end{equation}
in which $\lim_{\tau\rightarrow0}\tilde{N}(M^{D})=\tilde{N}_{IR}(\hat{M}_{IR}^{D})=0$
is used. 

The exponential of the counter term, $e^{\nu}<1$, which is usually
called the Gaussian density \citep{cao2004gaussian,cao2009recent}
in mathematics, is a relative volume or the reduced volume $\tilde{V}(M_{\tau=\infty}^{D})$
of the backwards limit manifolds introduced by Perelman. And $e^{\nu}$
is also the inverse of the initial condition of the manifolds density
$u_{\tau=0}^{-1}$. A finite value of the counter term $\nu(M_{\tau=\infty}^{D})$
makes an initial spacetime with unit fiducial volume at UV scale flow
and converge to a finite $u_{\tau=0}$ at IR. Thus the manifolds finally
converge to a finite relative volume instead of shrinking to a singular
point at IR $\tau=0$.

As an example, i.e. a homogeneous and isotropic late epoch universe,
with a positive curvature, its size in spatial and temporal parts
are on an equal footing (with a ``ball $B^{4}$'' radius $a(\tau)$)
, i.e. 
\begin{equation}
ds^{2}=a^{2}(\tau)\left(dX_{0}^{2}-dX_{1}^{2}-dX_{2}^{2}-dX_{3}^{2}\right),\label{eq:ball}
\end{equation}
which is nothing but a (Lorentzian) Shrinking Ricci Soliton configuration.
The solution tends to globally shrink $B^{4}$ to a singular point.
Note that the metric satisfies the shrinking soliton equation $R_{\mu\nu}=\frac{1}{2\tau}g_{\mu\nu}$,
and its volume form (\ref{eq:volume form}) are independent to the
signature, the counter term can be approximately given by a 4-ball
value $\nu(B_{\infty}^{4})\approx-0.8$ \citep{Luo:2019iby,Luo:2021zpi}.

Taken into account the counter term, the partition function (\ref{eq:relative-partition})
changes to
\begin{equation}
Z(M^{D})=e^{\lambda\tilde{N}-\frac{D}{2}-\nu(B_{\infty}^{4})},\label{eq:whole partition}
\end{equation}
which is anomaly canceled at UV and hence having a fixed fiducial
volume lab. 

Considering $\lim_{\tau\rightarrow0}\tilde{N}(M^{D})=0$ and $\lambda\int d^{D}Xu\tau\left|\nabla f_{\tau\rightarrow0}\right|^{2}=\frac{D}{2}$,
at low energy or small $\tau$, $\tilde{N}(M^{D})$ can be expanded
in powers of $\tau$
\begin{align}
\tilde{N}(M^{D}) & =\frac{\partial\tilde{N}}{\partial\tau}\tau+O(\tau^{2})=\tau\tilde{\mathcal{F}}+O(\tau^{2})\nonumber \\
 & =\int_{M^{D}}d^{D}Xu_{0}\left[\left(R_{\tau=0}+\left|\nabla f_{\tau=0}\right|^{2}-\frac{D}{2\tau}\right)\tau\right]+O(\tau^{2})\nonumber \\
 & =\int_{M^{D}}d^{D}Xu_{0}R_{0}\tau+O(\tau^{2}).
\end{align}
In this expansion, the effective action of $Z(M^{4})$ can be obtained
for $D=4$,
\begin{equation}
-\log Z(M^{4})=S_{eff}\approx\int_{M^{4}}d^{4}Xu_{0}\left(2\lambda-\lambda R_{0}\tau+\lambda\nu\right)\quad(\textrm{small\,\ensuremath{\tau}})
\end{equation}
Considering $u_{0}d^{4}X$ now recovers the classical invariant volume
element (\ref{eq:volume element}), $\sqrt{|g|}dV$, and by using
(\ref{eq:tau}) to change the flow time $\tau$ to the cutoff $k$,
we have
\begin{equation}
S_{eff}=\int_{M^{4}}dV\sqrt{|g|}\left(2\lambda-\frac{R_{0}}{64\pi^{2}}k^{2}+\lambda\nu\right)\quad(\textrm{small\,k}).\label{eq:eff-EH+cc}
\end{equation}

As the cutoff scale $k$ ranges from the lab scale to at least the
well-tested solar system scale ($k>0$), the action must recover the
Einstein-Hilbert (EH) action. However, at the cosmic scale ($k\rightarrow0$),
we know that the EH action deviates from observations and the cosmological
constant becomes important. Following the requirement, as $k\rightarrow0$,
the action leaving $2\lambda+\lambda\nu$ should play the role of
the standard EH action with a limit constant background scalar curvature
$R_{0}$ plus the cosmological constant $\Lambda$, so
\begin{equation}
2\lambda+\lambda\nu=\frac{R_{0}-2\Lambda}{16\pi G}.
\end{equation}
While at $k\rightarrow\infty$, $\lambda\tilde{N}\rightarrow\nu$,
the action leaving only the action $\frac{D}{2}\lambda=2\lambda$
for the fiducial lab, when it should be interpreted as a constant
EH action without the cosmological constant
\begin{equation}
2\lambda=\frac{R_{0}}{16\pi G}.
\end{equation}
So we have the cosmological term 
\begin{equation}
\lambda\nu=\frac{-2\Lambda}{16\pi G}=-\rho_{\Lambda}.
\end{equation}
As a result, the action can be rewritten as an effective EH action
plus a cosmological term
\begin{equation}
S_{eff}=\int_{M^{4}}dV\sqrt{|g|}\left(\frac{R_{k}}{16\pi G}+\lambda\nu\right)\quad(\textrm{small\,k}),
\end{equation}
where
\begin{equation}
\frac{R_{k}}{16\pi G}=2\lambda-\frac{R_{0}}{64\pi^{2}}k^{2},\label{eq:eff-R}
\end{equation}
which is in fact the solution of the flow equation of the scalar curvature
$\frac{\partial R}{\partial\tau}=-\frac{2}{D}R^{2}$, i.e. 
\begin{equation}
R_{k}=\frac{R_{0}}{1+\frac{1}{4\pi}Gk^{2}},\quad\textrm{or}\quad R_{\tau}=\frac{R_{0}}{1+\frac{2}{D}R_{0}\tau}.
\end{equation}
The flow equation of the scalar curvature is a homogeneous and isotropic
version of a more exact flow equation $\frac{\partial R}{\partial\tau}=-\Delta R-2R_{\mu\nu}R^{\mu\nu}$,
if we consider the scalar curvature is nearly homogeneous and isotropic
at IR i.e. $\Delta R=0$ and $R_{\mu\nu}=\frac{1}{D}Rg_{\mu\nu}$. 

Note that the effective scalar curvature is bounded by $R_{0}$ at
the cosmic scale $k\rightarrow0$, which can be measured by the ``Hubble's
constant'' $H_{0}$ at the cosmic scale,
\begin{equation}
R_{0}=D(D-1)H_{0}^{2}=12H_{0}^{2}.
\end{equation}
As a consequence, to recover the standard Einstein's gravity, $\lambda$
must be identical with the critical density of the universe
\begin{equation}
\lambda=\frac{3H_{0}^{2}}{8\pi G}=\rho_{c},\label{eq:critical density}
\end{equation}
so the cosmological constant here is predicted of order of the critical
density with a ``dark energy'' fraction
\begin{equation}
\Omega_{\Lambda}=\frac{\rho_{\Lambda}}{\rho_{c}}=-\nu\approx0.8,\label{eq:dark energy fraction}
\end{equation}
which is close to the observations. The detail discussions about the
cosmological constant problem and the cosmological effects, especially
the modification of the distance-redshift relation at the second order
by a deceleration parameter $q_{0}\approx-0.68$, can be found in
\citep{Luo:2019iby,Luo:2021zpi}. 

Finally, if matter is incorporated in the pure gravity action. $2\lambda$
term in eq.(\ref{eq:eff-(1)}) should be renormalized by the Ricci
flow, by using eq.(\ref{eq:eff-EH+cc}) and eq.(\ref{eq:eff-R}),
the effective action eq.(\ref{eq:eff-(1)}) at the 2nd order moment
or Gaussian approximation recovers the standard gravity+matter action
\begin{align}
S[\psi,X]\overset{(2)}{\approx} & \int dV\sqrt{|g|}\left[\frac{1}{2}g^{\mu\nu}\frac{\delta\psi}{\delta X^{\mu}}\frac{\delta\psi}{\delta X^{\nu}}-V(\psi)+2\lambda-\frac{R_{0}}{64\pi^{2}}k^{2}+\lambda\nu\right]\nonumber \\
= & \int dV\sqrt{|g|}\left[\frac{1}{2}g^{\mu\nu}\frac{\delta\psi}{\delta X^{\mu}}\frac{\delta\psi}{\delta X^{\nu}}-V(\psi)+\frac{R_{k}}{16\pi G}+\lambda\nu\right].
\end{align}

\section{The early Universe}

\subsection{Local Singularity Formation and Gradient Shrinking Ricci Soliton
Model}

In the previous section, by the cancellation effect of the cosmological
constant in the effective action of gravity, we have a non-singular
flow $t_{s}=0$ starting from a special metric (\ref{eq:ball}), i.e.
an initial spacetime where space and time are on an equal footing
at late epoch, homogeneous and with a positive curvature, however,
when starting from a more general initial spacetime manifolds, (e.g.
an early epoch Friedmann-Robertson-Walker metric (\ref{eq:FRW}) where
time is treated different from space), the Ricci (\ref{eq:ricci flow})
or the generalized Ricci-DeTurck flow (\ref{eq:ricci-deturk}) may
develop local curvature pinching or local singularities in some subset
of spacetime at finite scale $t_{s}\neq0$ \citep{1995Formation},
for instance, the early epoch singularity of the universe. The local
singularity can not be canceled as the global cosmological constant
did. 

For physical consideration, near the local singularity, the curvature
becomes larger and larger, the Ricci flow as the Gaussian approximation
of the RG-flow of NLSM may not be valid, because non-Gaussian terms
being composed of higher power of curvature may come into the flow
equation and become more and more important. However, it in fact depends
on how close the period producing our interested physics is to the
local singularity, we note that the observed primordial perturbations
produced from the early universe is highly Gaussian, while the non-Gaussian
parts are suppressed \citep{Planck:2018vyg}, which is a hint that
it is possible that the Ricci flow as a Gaussian approximation may
still be a good approximation in that period when the primordial perturbations
are produced, the non-Gaussian contributions to the RG-flow of the
NLSM could be safely ignored at least at the phenomenological level.

Thus if we still assume the validity of the Ricci flow when dealing
with the local curvature pinching, the no-local-collapsing theorem
\citep{perelman2002entropy,Luo:2022goc} of Perelman indicates the
local volume collapsing in the neighborhood of the singularity can
not actually occur at scale $O(\sqrt{t_{s}-t})$ and hence existing
a ``canonical neighborhood'' structure with a positive curvature
around the large curvature point at finite scale, the blow-up limit
of the canonical neighborhood around the local curvature pinching
point should resemble an ancient and self-similar configurations \citep{1993Eternal,2000Complete,2004Limiting}
satisfying the Gradient Shrinking Ricci Soliton (GSRS) equation. The
mathematical theorem provides us a possible way to study the early
universe where the curvature is highly pinched. In the following subsections
we will study the canonical and regular metric of the curvature pinched
region in the early universe and the primordial perturbations produced
from it. 

\subsection{Curvature Pinching of the FRW Metric at early Epoch: Inflation}

At sufficient large scale, if local irregularities (e.g. inhomogeneous,
shear \citep{Pitrou_2008} etc) of the universe are not large, the
Ricci flow gradually smooths out them making the universe seen spatially
more and more uniform as the cosmological principle asserts. So we
could always start from a spatially homogeneous isotropic Friedmann-Robertson-Walker
(FRW) metric $\mathbb{S}^{3}\times\mathbb{R}$
\begin{equation}
ds^{2}=dT^{2}-a^{2}(T,\tau)dX_{i}dX_{i},\label{eq:FRW}
\end{equation}
where the spatial metric, with $K$ the spatial curvature, i.e. $\textrm{\ensuremath{\textrm{Ric}^{(3)}}}=2Kg^{(3)}$,
has the form
\begin{equation}
dX_{i}dX_{i}=\frac{dr^{2}}{1-Kr^{2}}+r^{2}d\Omega^{2}.
\end{equation}
The Ricci-DeTurck flow (\ref{eq:ricci-deturk}) not only smooths out
the large scale universe to be FRW, but also tends to shrink the spatial
part $\mathbb{S}^{3}$ rather than the temporal part $\mathbb{R}$,
so it develops an early epoch singularity at certain singular scale
$t_{s}$. We consider the FRW ansartz at the vicinity of physical-time
origin $T=0$ (early epoch), and near the singular scale $t_{s}$,
i.e. $\tau=t_{s}-t\rightarrow0$, when the Ricci curvature pinches
to very large value, the spacetime local region around the singularity
is modeled by the GSRS equation (\ref{eq:shrinker}) as the flow limit
for the local spacetime with positive curvature \citep{perelman2002entropy,perelman2003ricci}
\begin{equation}
\textrm{\ensuremath{\textrm{Ric}^{(4)}}}+\nabla\nabla f=\frac{1}{2\tau}g^{(4)},\label{eq:GSRS equ-2}
\end{equation}
where $\textrm{Ric}^{(4)}$ is the Ricci curvature of 4-spacetime
\begin{equation}
\textrm{\ensuremath{\textrm{Ric}^{(4)}}}+\nabla\nabla f=\left(3\frac{\ddot{a}}{a}+\ddot{f}\right)dT^{2}+\left(2K-2\dot{a}^{2}-a\ddot{a}+a\dot{a}\dot{f}\right)dX_{i}dX_{i},
\end{equation}
and dot denotes the physical-time derivative $\dot{O}\equiv\frac{dO}{dT}$.
At leading order, to extend $a(T)$ and $f(T)$ smoothly to the origin
we naturally set the initial condition
\begin{equation}
\lim_{T\rightarrow0}a(T)\approx0,\quad\lim_{T\rightarrow0}\dot{f}(T)\approx0.\label{eq:T->0 limit}
\end{equation}
So we could first ignore the contribution from $f$ at leading order,
obtaining
\begin{equation}
3\ddot{a}-\frac{1}{2\tau}a=0,\label{eq:a-1}
\end{equation}
and
\begin{equation}
2\dot{a}^{2}+a\ddot{a}-2K-\frac{1}{2\tau}a^{2}=0.\label{eq:a-2}
\end{equation}
The solution of the (\ref{eq:a-1}) is an exponential expanding scale
factor, i.e. an inflation universe at early universe
\begin{equation}
a(T)=a(0)e^{H_{*}T},\label{eq:inflation a}
\end{equation}
where the Hubble rate is given by
\begin{equation}
H_{*}\equiv\frac{\dot{a}}{a}=\sqrt{\frac{1}{6\tau_{*}}}.\label{eq:H*}
\end{equation}
The subscript ``$*$'' in the following paper represents quantities
related to the particular physical-time and scale of the inflation
very near the singular scale $t_{s}$. At small $\tau_{*}\equiv t_{s}-t\rightarrow0$,
i.e. $t$ slightly deviating from $t_{s}$ ($\tau=0$), we have a
large expanding Hubble rate at early epoch. How close $\tau_{*}$
is to the singular limit $\tau=0$ are effectively described by the
input parameter: the e-folding number $\mathcal{N}$, see next subsection.

As $a(T)$ is exponentially expanding, the spatial curvature term
in (\ref{eq:a-2}) becomes more and more unimportant, the observed
universe is seen more and more spatially flat, then the solution of
(\ref{eq:a-2}) finally also tends to the exponential expanding profile
(\ref{eq:inflation a}) as the common inflation solution of (\ref{eq:a-1})
and (\ref{eq:a-2}).

In the framework, first, we see that a deSitter like configuration
(\ref{eq:GSRS equ-2}) (up to a diffeomorphism (\ref{eq:R-ddlogu}))
and hence a spatially inflationary universe is naturally obtained
from the Ricci flow limit, so a Bunch-Davies vacuum as a common choice
of vacuum state of inflation and nearly scale-invariant spectrum are
naturally obtained at the leading order. Second, we note that the
inflation behavior at early epoch is not driven by any speculative
inflaton fields, but caused by a completely different mechanism, i.e.
the shrinking behavior of the GSRS (\ref{eq:shrinker}) or (\ref{eq:GSRS equ-2}),
which is also closely related to the conformal instability of the
system. The above simple result shows, at the classical solution level,
the early epoch inflation is a natural consequence when the spacetime
is driven by the Ricci flow to form local singularity at the vicinity
of the physical-time origin. 

\subsection{Small Deviation from Exact deSitter: Slow Roll Parameters}

The small deviation of $\tau$ from zero not only gives a large and
finite Hubble rate $H_{*}$ at early epoch, but also causes a small
deviation from the exact deSitter configuration. Here we consider
the scale factor $a(T,0)$ at $\tau=0$ is an exact deSitter configuration,
i.e. the Hubble rate $H_{\tau=0}$ is exactly a constant independent
to the physical-time $T$, and we calculate the physical-time dependence
of the Hubble rate $H(T,\tau_{*})$ when $\tau_{*}$ slightly deviates
from exact 0. Since $\tau_{*}$ is a small, we can expand the scale
factor in powers of $\tau$,
\begin{equation}
a(T,\tau_{*})=a(T,0)\left(1+b_{1}\tau_{*}+...\right),\quad(\tau_{*}\rightarrow0,\:T\rightarrow0).
\end{equation}
Substituting it into the Ricci flow equation (\ref{eq:ricci flow}),
\begin{equation}
\frac{\partial a^{2}(T,\tau)}{\partial\tau}=2\left(2\dot{a}^{2}+a\ddot{a}\right).
\end{equation}

By using $\frac{\dot{a}}{a}=H_{\tau=0}$ and $\dot{H}_{\tau=0}\ll H_{\tau=0}^{2}$,
we have the expansion coefficient
\begin{equation}
b_{1}=3H_{\tau=0}^{2},
\end{equation}
so the Ricci flow gives rise to a rescaling to the scale factor, which
can also approximately interpreted as a physical-time varying of the
Hubble rate, meaning a deviation from exact deSitter, i.e.
\begin{equation}
a(T,\tau_{*})=a(T=0)e^{H_{\tau=0}T}\left(1+3H_{\tau=0}^{2}\tau_{*}+...\right)\approx a(T=0)e^{H_{*}T},
\end{equation}
where
\begin{equation}
H_{*}(T)\equiv H_{*}(T,\tau_{*})\approx H_{\tau=0}+\frac{3H_{\tau=0}^{2}\tau_{*}}{T}.\label{eq:H*(T)}
\end{equation}

In the standard terminology of inflation, the ``slow roll parameter''
as an approximate constant parameter is given by the physical-time
derivative of the Hubble rate during the inflation, but in the case
(\ref{eq:H*(T)}), the slow roll parameter is also change with $T$,
so a typical-time $T_{*}$ is needed to approximately gives a frozen
value or typical value of the slow roll parameter 
\begin{equation}
\epsilon_{*}\equiv\epsilon(T_{*})\equiv-\frac{\dot{H}_{*}(T_{*})}{H_{*}^{2}(T_{*})}=\frac{3H_{\tau=0}^{2}\tau_{*}}{H_{*}^{2}T_{*}^{2}},\label{eq:epsilon}
\end{equation}
where $T_{*}=\gamma^{-1}T_{end}<T_{end}$ is a typical-time of the
inflation when the Hubble rate can be considered as constant $H_{\tau=0}\approx H_{*}$
and hence we could takes $\epsilon(T_{*})$ as its typical value during
the inflation. The typical-time $T_{*}$ is expected several times
earlier than the end-time $T_{end}$ of the inflation, for instance,
\begin{equation}
\gamma=\frac{T_{end}}{T_{*}}>O(1).\label{eq:gamma}
\end{equation}

Therefore, if we consider the Hubble rate $H_{\tau=0}\approx H_{*}=\frac{1}{\sqrt{6\tau_{*}}}$
and the typical-time $T_{*}$ are both constants during the inflation,
then we have a constant slow roll parameter during the inflation 
\begin{equation}
\epsilon_{*}\simeq\frac{\gamma^{2}}{2\mathcal{N}^{2}},\label{eq:epsilon-N2}
\end{equation}
where the e-folding number $\mathcal{N}$ is defined as 
\begin{equation}
\mathcal{N}=\ln\frac{a(T_{end},\tau_{*})}{a(0,0)}\approx H_{*}T_{end}\gg1.
\end{equation}
The condition $|\epsilon_{*}|\ll1$ is called the slow roll approximation,
resulting $\sqrt{\tau_{*}}\ll T_{*}<T_{end}$, leading to a small
deviation of the inflation background from deSitter.

In the framework, we also note that the term \textquotedblleft slow
roll parameters\textquotedblright{} is just for historical convention,
it does not relate to any scalar fields rolling down a potential.
The small but finite value of the parameter are completely due to
a small deviation of $\tau_{*}$ from the singular flow-time $\tau=0$
governed by short flow-time evolution of the Ricci flow. 

\subsection{End of the Inflation}

The GSRS configuration mimics the local spacetime near the curvature
pinching point $T=0$. As the physical-time $T$ evolves or varies
away from the local point $T=0$, the local GSRS part should smoothly
connect to the non-deSitter rest part of the spacetime manifolds.
These two parts are connected at about $T_{end}$ when the inflation
comes to an end, satisfying $\epsilon(T_{end})\simeq1$. If we consider
that $H_{\tau=0}$ in the local GSRS part is much larger than $H_{end}$
in the non-deSitter rest part, as $H_{*}$ in (\ref{eq:epsilon})
gradually decreases to $H_{end}$, we have $|\epsilon_{*}|\ll1$ gradually
increase to be of order one,
\begin{equation}
\epsilon(T_{end})\simeq\frac{3H_{\tau=0}^{2}\tau_{*}}{H_{end}^{2}T_{end}^{2}}\simeq1.
\end{equation}
 It leads to the estimate that $T_{end}$ is much larger than $\sqrt{\tau_{*}}\sim O(H_{*}^{-1})$,
\begin{equation}
T_{end}\simeq\frac{H_{\tau=0}}{H_{end}}\sqrt{\tau_{*}},
\end{equation}
which is much longer than $H_{*}^{-1}$ consistent with general expectations
in the standard inflation theories. 

We note that since $T_{*}$ and $T_{end}$ are both much larger than
the local non-collapsing scale $O(\sqrt{t_{s}-t_{*}}=\sqrt{\tau_{*}})$
of the Ricci flow. Thus the ``canonical neighborhood'' structure
is indeed well-defined during the inflation, which is the self-consistent
reason why we could model the inflation period in the early universe
by a GSRS metric.

\subsection{Primordial Perturbations}

We have proved that at leading order the Ricci flow limit (GSRS) gives
rise to an inflationary universe at early epoch. Important phenomenology
of the early universe come from the primordial perturbations at the
next leading order. In this section, we consider the scalar and tensor
primordial perturbations produced during the inflation period. 

Note that there are two kinds of scalar perturbations in the theory,
first is the scalar (Newtonian potential) perturbation $\varphi$
around the inflation background,
\begin{equation}
ds^{2}=\left(1+2\varphi\right)dT^{2}-a^{2}(T,\tau_{*})\left(1-2\varphi\right)dx_{i}dx_{i},\quad(0<\sqrt{\tau_{*}}<T<T_{end})
\end{equation}
and the second is the scalar perturbation $\delta u$ around the density
$u_{*}(T)\equiv u(T,\tau_{*})$ 
\begin{equation}
u(\mathbf{k},T)=u_{*}(T)+\delta u(\mathbf{k},T),
\end{equation}
where $\mathbf{k}$ is the Fourier modes of the fluctuation. In fact,
the additional scalar ``field'' $u$ appearing in the effective
action (\ref{eq:I0+I2}) is inevitable, and plays a fundamental role
in the framework, already introduced in the subsection D of section
II. It not only leads to the conformal instability and hence inflation
due to the \textquotedbl wrong sign\textquotedbl{} in its kinetic
term (see later), but also has direct statistic and geometric meanings
\citep{Luo:2022statistics}.

These two kinds of scalar fluctuations can be mixed up into a gauge
invariant scalar perturbation
\begin{equation}
\delta\phi\equiv-\delta u+\frac{\dot{u}_{*}}{H_{*}}\varphi.
\end{equation}
By using the new variable $\delta\phi$, the fixed point action $\tilde{N}$
in (\ref{eq:whole partition}) (up to constants) can be expanded to
the quadratic order of $\delta\phi$,
\begin{align}
\tilde{N} & =\tau_{*}\tilde{\mathcal{F}}+...=\lambda\tau_{*}\int_{M^{D}}d^{D}X\left[uR+\frac{1}{u}|\nabla u|^{2}-\frac{D}{2\tau}u\right]=\tau_{*}\left[I_{0}(H_{*},u_{*})+I_{2}(\delta\phi)+...\right],\quad(\tau_{*}\rightarrow0).\label{eq:I0+I2}
\end{align}

The action seems belong to a wide class of scalar-tensor theory of
gravity \citep{2003The} with the density $u$ playing the role of
a ``scalar field''. Note that the kinetic term of $u$ has a ``wrong
sign'' which gives rise to a conformal instability \citep{Luo:2022goc}
to the gravitational system. Such instability is the essential reason
for the singularity formation and inflation. Since $u$ can also be
seen as a ``conformal factor'' of gravity, in a proper gauge, the
instability of $u$ can also be transformed to and interpreted as
the instability of the metric leading to the inflation background
eq.(\ref{eq:inflation a}), leaving $u_{*}$ does not directly feel
the instability and mildly changes under the gauge. More precisely,
note $R_{*}=6(2H_{*}^{2}+\dot{H}_{*})$ being the scalar curvature
in inflation, the lowest order fixed point action is given by
\begin{equation}
I_{0}(H_{*},u_{*})=\lambda\int_{M^{4}}d^{3}XdT\frac{1}{u_{*}}\left\{ \dot{u}_{*}^{2}+\left[6\left(2H_{*}^{2}+\dot{H}_{*}\right)-\frac{2}{\tau_{*}}\right]u_{*}^{2}\right\} ,\label{eq:i0}
\end{equation}
and by using (\ref{eq:H*}) and $1/u_{*}=a^{3}$, its Euler-Lagrangian
equation for $u_{*}$ gives rise to the conjugate heat equation (\ref{eq:u-equation})
taking the form
\begin{equation}
\ddot{u}_{*}+3H_{*}\dot{u}_{*}-6\dot{H}_{*}u_{*}=0,\label{eq:classical u eq}
\end{equation}
in which $u_{*}$ does not feel instability and mildly changes under
the gauge. If $\ddot{u}_{*}$ compared with other terms can be ignored
then we have
\begin{equation}
\frac{\dot{u}_{*}}{u_{*}}=\frac{2\dot{H}_{*}}{H_{*}}=-2\epsilon_{*}H_{*}.\label{eq:u'/u}
\end{equation}

$I_{0}=0$ as the fixed point equation of the gradient flow of $\tilde{N}$
also gives rise to the classical GSRS equation (\ref{eq:GSRS equ-2})
and hence gives the deSitter spacetime (\ref{eq:inflation a}) near
$T=0$. By considering $\dot{H}_{*}\sim O(\epsilon_{*}H_{*}^{2})$
and $\dot{u}_{*}/u_{*}\sim O(\epsilon_{*}H_{*})$ proportional to
$\epsilon_{*}$ are both small, so at leading order, extremizing $I_{0}(H_{*},u_{*})=0$
recovers the result (\ref{eq:H*}).

At the next leading order, the fixed point action at the quadratic
order of $\delta\phi$ is \citep{Jai1996Unified,Hwang:1996bc,DeFelice:2010aj}
\begin{equation}
I_{2}(\delta\phi)=\frac{1}{2}\lambda\int_{M^{4}}d^{3}XdTZ\left\{ \delta\dot{\phi}^{2}-\frac{1}{a^{2}}|\nabla\delta\phi|^{2}+\frac{1}{a^{3}Z}\frac{H_{*}}{\dot{u}_{*}}\left[a^{3}Z\left(\frac{\dot{u}_{*}}{H_{*}}\right)^{\centerdot}\right]^{\centerdot}\delta\phi^{2}\right\} 
\end{equation}
with 
\begin{equation}
Z=\frac{1}{u_{*}}\frac{1}{\left(1+\frac{1}{2H_{*}}\frac{\dot{u}_{*}}{u_{*}}\right)^{2}}\approx\frac{1}{u_{*}}(1+2\epsilon_{*}),
\end{equation}
where (\ref{eq:u'/u}) has been used. The action is the starting point
for studying the spectrum of the primordial scalar perturbation on
the inflationary background given by $I_{0}(H_{*},u_{*})$.

When $u_{*}=1$ and $\epsilon_{*}=0$, we have $Z=1$, then $I_{2}(\delta\phi)$
recovers the action of standard minimally coupled scalar inflation
action that induces the standard Mukhanov-Sasaki equation. In this
theory, the Euler-Lagrangian equation of $I_{2}(\delta\phi)$ gives
rise to a Z-modified Mukhanov-Sasaki equation
\begin{equation}
\delta\ddot{\phi}+\frac{(a^{3}Z)^{\cdot}}{a^{3}Z}\delta\dot{\phi}+\left\{ \frac{\mathbf{k}^{2}}{a^{2}}-\frac{1}{a^{3}Z}\frac{H_{*}}{\dot{u}_{*}}\left[a^{3}Z\left(\frac{\dot{u}_{*}}{H_{*}}\right)^{\centerdot}\right]^{\centerdot}\right\} \delta\phi=0.
\end{equation}
Similar with the standard treatment, we introduce
\begin{equation}
v(\mathbf{k},T)=\sqrt{Z}a\delta\phi,\quad z=\frac{a}{H_{*}}\sqrt{Z\dot{u}_{*}^{2}},
\end{equation}
then the Z-modified Mukhanov-Sasaki equation becomes \citep{DeFelice:2010aj}
\begin{equation}
v^{\prime\prime}+\left(\mathbf{k}^{2}-\frac{z^{\prime\prime}}{z}\right)v=0,
\end{equation}
in which prime denotes the conformal physical-time $d\eta=a^{-1}dT$
derivative.

By standard consideration, in the subhorizon limit $\mathbf{k}^{2}\gg\frac{z^{\prime\prime}}{z}$,
the $v$ is fast oscillating inside the horizon, while in the superhorizon
limit, $\mathbf{k}^{2}\ll\frac{z^{\prime\prime}}{z}$, the solution
can be written as a Hankel functions $H_{\nu}^{(1)}$ in terms of
the conformal physical-time $\eta$ \citep{DeFelice:2010aj},
\begin{equation}
v_{\mathbf{k}}(\eta)=\frac{\sqrt{\pi|\eta|}}{2}e^{i(1+2\nu)\pi/4}H_{\nu}^{(1)}(\mathbf{k}|\eta|),
\end{equation}
where 
\[
\nu^{2}=\frac{1}{4}+\frac{z^{\prime\prime}}{z}\eta^{2}.
\]
If the slow roll parameters are considered constant (not vary with
conformal physical-time), then we have
\begin{equation}
\nu^{2}=\frac{1}{4}+\frac{(1+\epsilon_{*}+\delta_{*}+\alpha_{*})(2+\delta_{*}+\alpha_{*})}{(1-\epsilon_{*})^{2}},
\end{equation}
where the slow roll parameters are defined as
\begin{equation}
\epsilon_{*}\equiv-\frac{\dot{H}_{*}}{H_{*}^{2}},\quad\delta_{*}\equiv-\frac{\ddot{u}_{*}}{H_{*}\dot{u}_{*}},\quad\alpha_{*}\equiv-\frac{\dot{u}_{*}}{2H_{*}u_{*}}=\epsilon_{*},\label{eq:slow roll para}
\end{equation}
in which (\ref{eq:u'/u}) have been used in the 3rd parameter.

Thus the scalar power spectrum is
\begin{equation}
P_{\delta\phi}\equiv16\pi G\frac{\mathbf{k}^{3}}{2\pi^{2}}|\delta\phi_{\mathbf{k}}|^{2},
\end{equation}
where $G$ is Newton's constant. We finally have
\begin{equation}
P_{\delta\phi}=\frac{16\pi G}{Q}\left[(1-\epsilon_{*})\frac{\Gamma(\nu)}{\Gamma(3/2)}\frac{H_{*}}{2\pi}\right]^{2}\left(\frac{|\mathbf{k}\eta|}{2}\right)^{3-2\nu},
\end{equation}
where $\Gamma$ is the Gamma function. The power spectrum of scalar
perturbation is frozen when they cross to the outside of the horizon
for which $\mathbf{k}=aH_{*}$, in the slow roll approximation the
power spectrum of the scalar perturbation can be given by \citep{DeFelice:2010aj}
\begin{equation}
P_{\delta\phi}=\frac{16\pi G}{Q}\left.\left(\frac{H_{*}}{2\pi}\right)^{2}\right|_{\mathbf{k}=aH_{*}}\approx\frac{4\pi G}{\epsilon_{*}^{2}u_{*}}\left.\left(\frac{H_{*}}{2\pi}\right)^{2}\right|_{\mathbf{k}=aH_{*}},\label{eq:Ps}
\end{equation}
where
\begin{equation}
Q=Z\left(\frac{\dot{u}_{*}}{H_{*}}\right)^{2}\approx\frac{1}{u_{*}}(1+2\epsilon_{*})\left(\frac{\dot{u}_{*}}{H_{*}}\right)^{2}\approx4\epsilon_{*}^{2}u_{*}.
\end{equation}
The spectral index for the scalar perturbation is given by
\begin{equation}
n_{s}-1\equiv\frac{d\ln P_{\delta\phi}}{d\ln\mathbf{k}}\approx-4\epsilon_{*}-2\delta_{*}-2\alpha_{*}.\label{eq:ns-1}
\end{equation}

Tensor perturbations $h_{ij}$ in $g_{ij}=a^{2}(T,\tau_{*})(\delta_{ij}+h_{ij})$
have two polarization states $h_{p}$ where $p=+,\times$. By using
the polarization tensors bases $e_{ij}^{+}$ and $e_{ij}^{\times}$,
we have
\begin{equation}
h_{ij}=h_{+}e_{ij}^{+}+h_{\times}e_{ij}^{\times}.
\end{equation}
For a similar consideration followed by the scalar perturbations,
and in the slow roll approximation, the power spectrum of each polarization
component $h_{p}$ ($p=+,\times$) is given by \citep{DeFelice:2010aj}
\begin{equation}
P_{h}\approx\frac{128\pi G}{u_{*}}\left.\left(\frac{H_{*}}{2\pi}\right)^{2}\right|_{\mathbf{k}=aH_{*}},\label{eq:Ph}
\end{equation}
with a much smaller spectral index beyond the order of slow roll approximation,
\begin{equation}
n_{t}\equiv\frac{d\ln P_{h}}{d\ln\mathbf{k}}\approx-2\epsilon_{*}+2\alpha_{*}=-2\epsilon_{*}+2\epsilon_{*}=0.
\end{equation}
The power spectrum of the tensor perturbation is much smaller and
more scale-invariant than the prediction of the standard inflation.

\subsection{Estimate of $u_{*}$ and $H_{*}$}

To predict the power spectrums of the scalar and tensor perturbations
(\ref{eq:Ps}) (\ref{eq:ns-1}) and (\ref{eq:Ph}) more precisely,
several parameters are needed to further estimate in this framework,
the most important ones are the $u_{*}\equiv u(T,\tau_{*})$ and $H_{*}\equiv H(T_{*},\tau_{*})$.

Since $u_{*}$ represents the volume ratio (\ref{eq:coarse-grain density})
between the fiducial 3-volume (standard scale factor $a=1$) and the
local 3-volume of early universe, by using (\ref{eq:volume constraint})
and (\ref{eq:inflation a}), the physical-time derivative of $u$
is volume changing rate during the inflation is
\begin{equation}
\lim_{T\rightarrow0}\dot{u}_{*}=-3H_{*}.
\end{equation}
By using (\ref{eq:u'/u}), so we have
\begin{equation}
\lim_{T\rightarrow0}u_{*}=\frac{3}{2\epsilon_{*}}\sim\frac{1}{a^{3}(T_{*},\tau_{*})},\label{eq:u0}
\end{equation}
so
\begin{equation}
u_{*}(T)=\frac{3}{2\epsilon_{*}}-3H_{*}T\approx\frac{3}{2\epsilon_{*}}e^{-2\epsilon_{*}H_{*}T},\quad(T\rightarrow0),
\end{equation}
which is obviously also an approximate solution of (\ref{eq:classical u eq}).
The estimate of $u_{*}$ make small corrections of order $O(\epsilon_{*})$
to the unperturbed conditions (\ref{eq:T->0 limit}) at the slow roll
approximation.

From the estimate of $u_{*}$, we can directly have the second slow
roll parameter $\delta_{*}$ in (\ref{eq:slow roll para})
\begin{equation}
\delta_{*}\equiv-\frac{\ddot{u}_{*}}{H_{*}\dot{u}_{*}}=-\frac{\dot{H}_{*}}{H_{*}^{2}}=\epsilon_{*}.
\end{equation}
In this situation, if we take $\mathcal{N}\simeq60$, the spectral
index of the scalar perturbation (\ref{eq:ns-1}) is in the range
\begin{equation}
n_{s}=1-8\epsilon_{*}\simeq1-\frac{4\gamma^{2}}{\mathcal{N}^{2}}\simeq(0.91\sim0.99),\label{eq:ns}
\end{equation}
if taking $1<\gamma\apprle9$, describing how close the typical-time
$T_{*}$ to the end-time $T_{end}$ of inflation. The favored value
of observations $n_{s}\simeq0.96$ can be obtained if one takes $\gamma\approx6$,
which is consistent with the pre-assumption (\ref{eq:gamma}).

In terms of Perelman's seminal introduction of his monotonic functionals
and reduced volume, the Ricci flow spacetime admits a remarkable comparison
geometry picture. From the geometric point of view, one of the most
crucial information of the geometry is coded in the local volume comparison.
Remind in the subsection II-F that the fraction of the ``dark energy''
w.r.t. the critical density is related to the relative volume or reduced
volume $u_{\tau=0}^{-1}=\tilde{V}(M_{\tau=\infty}^{D})=e^{\nu}<1$
of order one, and the estimate of $H_{*}$ in the local curvature
pinching region also belongs to such kind of question in the framework.
The value of $H_{*}\equiv H(T_{*},\tau_{*})$ not only depends on
the typical-time $T_{*}$, but also on the scale $\tau_{*}$, and
its rough order of magnitude at fixed $T_{*}$ is almost given by
$\tau_{*}$ and hence the e-folding $\mathcal{N}$. The theory has
the critical density $\lambda=\rho_{c}\sim(10^{-3}eV)^{4}$ (\ref{eq:critical density})
as the only dimensional input of the theory, and together with the
scale factor $a(0)\sim e^{-\mathcal{N}}$. So by dilating the current
critical energy scale $\lambda^{1/4}$ by the scale factor $a(0)$
of the inflation period, we have a natural estimate to the energy
scale and Hubble rate at the typical-time of the inflation if taking
$60\apprle\mathcal{N}\apprle70$
\begin{equation}
H_{*}\simeq\lambda^{1/4}a^{-1}(0)\simeq(10^{14}\sim10^{18})\textrm{GeV},\label{eq:H* value}
\end{equation}
which is within a generally accepted range of inflation scale. The
estimate of $H_{*}$ near the local singularity has direct relation
to the estimate (e.g. Harnack estimate) of a local curvature and hence
the local volume comparison in the Ricci flow. This estimate also
suggests a possible picture for the mysterious large orders of magnitude
between the energy scales of the early epoch accelerating inflation
(of order of $H_{*}$) and the late epoch accelerating expansion (corresponding
to the cosmological constant of order of $\lambda^{1/4}$): the high
energy scale of $H_{*}$ is because the local pinching curvature is
almost unbounded given by the large e-folding number $a^{-1}(0)\sim e^{\mathcal{N}}\gg1$
at the early epoch, while the cosmological constant is of order of
$\lambda$ at late epoch, with fraction of order one, i.e. (\ref{eq:dark energy fraction}). 

By using (\ref{eq:u0}), (\ref{eq:H* value}) and (\ref{eq:epsilon-N2})
the scalar power spectrum (\ref{eq:Ps})
\begin{equation}
P_{\delta\phi}\approx\frac{4\pi G}{\epsilon_{*}^{2}u_{*}}\left.\left(\frac{H_{*}}{2\pi}\right)^{2}\right|_{\mathbf{k}=aH_{*}}=\frac{8\pi G}{3\epsilon_{*}}\left.\left(\frac{H_{*}}{2\pi}\right)^{2}\right|_{\mathbf{k}=aH_{*}}\simeq10^{-5}
\end{equation}
is within the range of observations when taking $\mathcal{N}\simeq60$
with $\gamma\simeq6$. 

The tensor power spectrum (\ref{eq:Ph}) is predicted as
\begin{equation}
P_{h}\approx\frac{128\pi G}{u_{*}}\left.\left(\frac{H_{*}}{2\pi}\right)^{2}\right|_{\mathbf{k}=aH_{*}}=\frac{256\pi G}{3}\epsilon_{*}\left.\left(\frac{H_{*}}{2\pi}\right)^{2}\right|_{\mathbf{k}=aH_{*}}\simeq10^{-8},
\end{equation}
which is more difficult to be detected than the standard prediction.
And the tensor-scalar ratio is given by
\begin{equation}
r\equiv\frac{P_{h}}{P_{\delta\phi}}\simeq32\epsilon_{*}^{2}\simeq\frac{8\gamma^{4}}{\mathcal{N}^{4}}\simeq0.0008,\label{eq:r}
\end{equation}
when taking $\mathcal{N}\simeq60$ with $\gamma\simeq6$, which compared
with the standard prediction is more difficult to be observed and
hence is certainly consistent with current observations. 

And similar with standard consideration, (\ref{eq:ns}) and (\ref{eq:r})
can be combined to give a relation, 
\begin{equation}
r=\frac{1}{2}(1-n_{s})^{2}.
\end{equation}
This relation implies that the prediction in $r-n_{s}$ plane is independent
of the parameters $\mathcal{N}$ and $\gamma$, which is inside the
allowed range of the observations.

\section{Summary and Conclusions}

The paper reviews the quantum fields theory of spacetime reference
frame. The theory is described by a non-linear sigma model in $d=4-\epsilon$
base space, and the target space is interpreted as the quantum reference
frame fields. The 2nd order central moment quantum fluctuation of
the frame fields introduces the Ricci flow to the spacetime manifolds
$(M,g)$ at Gaussian approximation. The normalized density matrix
$u$ of the theory can also be written explicitly at the Gaussian
approximation, which induces the Ricci-DeTurck flow to the spacetime
manifolds with density $(M,g,u)$. We use the functional integral
method to deduce the diffeomorphism anomaly and the effective action
of gravity based on the quantum spacetime theory. 

When we apply the Ricci flow to the late epoch of the universe when
the space and time are on an equal footing (\ref{eq:ball}), $B^{4}$,
homogeneous and with positive curvature, the Ricci flow globally shrinks
the spacetime $B^{4}$ isotropically. Since the base space interpreted
as the fiducial lab is considered rigid, the diffeomorphism anomaly
must be canceled in the fiducial lab up to UV scale, as a consequence,
a cosmological constant emerging into the effective action can be
calculated by the anomaly cancellation, which also normalizes the
spacetime and prevents the Ricci flow shrinking the spacetime into
a singular point but to a limit spacetime with finite relative volume
(w.r.t. the fiducial lab volume). 

When applying the Ricci flow to the early universe when time and space
are treated differently (\ref{eq:FRW}), $\mathbb{S}^{3}\times\mathbb{R}$,
the Ricci flow locally shrinks $\mathbb{S}^{3}$ rather than $\mathbb{R}$,
and hence develops local curvature pinching near the origin of the
physical-time. In this case, the local singularity can no longer be
``normalized out'' as the global cosmological constant does in the
case of equal-footing-spacetime. The closeness of the early epoch
producing the primordial perturbations to the singularity is effectively
described by a finite e-folding number $\mathcal{N}$, due to the
highly suppressed non-Gaussian primordial fluctuations in the cosmic
observations, if we still assume the validity of the Ricci flow (being
a Gaussian approximate) applying to the early epoch, the high curvature
region of a manifold resembles a rescaled ancient non-collapse solution,
the local non-collapsing theorem and the Gradient Shrinking Ricci
Soliton (GSRS) model of the local singularity provide us possible
approach to the quantum treatment of the early universe. 

The paper shows that the Ricci flow limit configuration (GSRS) resembling
the large curvature region of the spacetime is a promising model for
the early universe because the following results are obtained: (i)
the Ricci flow deforms a spatial homogeneous and isotropic FRW ansartz
to the GSRS flow limit, which mimics a spatially inflationary (deSitter)
universe at the leading order; (ii) the self-similarity of the GSRS
configuration provides a natural explanation of nearly scale invariant
and Gaussian spectrum of the primordial perturbations in cosmic observations;
(iii) a finite scale $\tau_{*}$ slightly deviating from the singular
flow-time $\tau=0$ not only at leading order gives rise to a large
inflation Hubble rate $H_{*}$ but also at the next leading order
gives a small deviation from exact deSitter and hence gives rise to
small slow roll parameters and small deviation of the primordial perturbation
spectrum from exact scale invariant. 

The slow roll parameters and the related inflation ending are also
discussed within the framework. Because during the inflation, the
geometric quantities such as the scale factor $a(T,\tau)$, the Hubble
rate $H(T,\tau)$, and the manifolds density $u(T,\tau)$ are functions
of both the flow-time $\tau$ and the physical-time $T$, so we need
to further estimate the scale $\tau_{*}$ and typical-time $T_{*}$
of the inflation, to evaluate their typical values during the inflation.
It leads to 3 inputs in the calculating of the slow roll parameters
and the power spectrum of the primordial perturbations: (a) the e-folding
number $\mathcal{N}$ describing how close the scale $\tau_{*}$ of
the inflation producing our interested physics is to the singular
scale $\tau=0$, (b) $\gamma=T_{end}/T_{*}$ describing how close
between the end-time and the typical-time of the inflation, when the
typical values of the slow roll parameters are taken and the Hubble
rate $H_{*}$ can be considered as a constant, and beside those we
also have (c) $\lambda=\rho_{c}\simeq(10^{-3}eV)^{4}$ being the fundamental
dimensional input of the frame fields theory. $\lambda$ is fixed
from the first principle of the theory, while $\mathcal{N}$ and $\gamma$
can be tuned in some ranges: e.g. the e-folding is taken within $60\apprle\mathcal{N}\apprle70$,
and $\gamma$ as the ratio between $T_{end}$ and $T_{*}$ must only
be several times larger than 1. If we take $\mathcal{N}\simeq60$
with $\gamma\simeq6$, first, the energy scale of the local curvature
pinching of the early epoch (\ref{eq:H* value}) is estimated consistent
with generally accepted range of the standard inflation scale; and
second, the predicted power spectrum and index of the primordial scalar
perturbation could be consistent with observations, and the ones of
the tensor perturbation are much smaller than the standard inflation
theory, which are even more difficult to be detectable. Thus it seems
that the tensor perturbation is too small to be used to distinguish
this theory and the standard inflation theory. 

The Ricci flow is a very powerful tool to study the geometry especially
its short distance as well as long distance structure, and provide
us a framework to ask meaningful questions to the geometry and physics.
So if the approach really hits certain core of the geometry and physics
in the early universe, we hope, beside the observables studied in
the paper (taking the standard inflationary paradigm as the reference),
our future investigations of the theory will provide us more possible
predictions and insights to the early universe. 
\begin{acknowledgments}
This work was supported in part by the National Science Foundation
of China (NSFC) under Grant No.11205149, and the Scientific Research
Foundation of Jiangsu University for Young Scholars under Grant No.15JDG153.

\bibliographystyle{plain}

\end{acknowledgments}

\end{document}